\documentstyle[12pt]{amsart} \mathsurround=2pt
\newtheorem{th}{Theorem}[section]
\newtheorem{crl}[th]{Corollary}
\newtheorem{prp}[th]{Proposition}
\newtheorem{lm}[th]{Lemma}

\newcommand{\der}{\partial}
\newcommand{\LL}{{\mathcal L}}
\newcommand{\WW}{{W}}

\newcommand{\eps}{\varepsilon}

\begin{document}

\title{$10$-commutator and $13$-commutator }
\author{A.S. Dzhumadil'daev}
\address
{Kazakh-British University, Tole bi, 59, Almaty, 480091,
Kazakhstan} \email{askar@@math.kz} \maketitle

\bigskip

\begin{abstract}
Skew-symmetric sum of $N!$ compositions of $N$ vector fields in
all possible order is called $N$-commutator. We construct
$10$-commutator and $13$-commutator on $Vect(3)$ and
$10$-commutator on a space of divergenceless vector fields
$Vect_0(3).$ We show that $2$-commutator, $10$-commutator and
$13$-commutator  form final list of $N$-commutators on $Vect(3)$
and under these polylinear operations $Vect(3)$ has a structure of
sh-Lie algebra. We establish that the list of $2$-and
$10$-commutators on $Vect_0(3)$ is also final. Constructions are
based on calculations of powers of odd derivations.
\end{abstract}

Let $(A,\circ)$ be an algebra with vector space $A$ and
multiplication $\circ$. Let ${\bf C}<t_1,\ldots,t_k>$ be a space
of non-commutative non-associative polynomials. Any $f\in {\bf
C}<t_1,\ldots,t_k>$ induces a $k$-ary map $$f:
\underbrace{A\times\cdots \times A}_k\rightarrow A,$$ that
corresponds to any $a_1,\ldots,a_k\in A$  element
$f(a_1,\ldots,a_k)$ calculated in terms of multiplication $\circ.$
If this
map is trivial, i.e., $f(a_1,\ldots,a_k)=0,$ for any
$a_1,\ldots,a_k\in A$ then $f=0$ is said an {\it identity} on
$(A,\circ).$ If $f$ is polylinear, then $f$ induces a {\it $k$-ary
multiplication} on $A.$ For example, if $s_2=t_1t_2-t_2t_1\in {\bf
C}<t_1,t_2>,$ then $$s_2(a,b)=a\circ b-b\circ a$$ is an ordinary
commutator.

Let $$s_k=\sum_{\sigma\in
Sym_k}sign\,\sigma\,t_{\sigma(1)}(\cdots(t_{\sigma(k-1)}t_{\sigma(k)})\cdots)$$
be standard skew-symmetric polynomial. Let $Dif\!f_n$ be a space
of differential operators with $n$ variables. For simplicity
assume that variables are from ${ \bf C}[x_1,\ldots,x_n].$ Let
$Dif\!f_n^{[d]}$ be a subspace of differential operators of order
$d$: $$Dif\!f_n^{[d]}=\left< u\der^{\alpha} |
|\alpha|=\sum_{i=1}^n\alpha_i=d\right>.$$ We can interpret
differential operators of first order as  vector fields and
identify $Dif\!f_n^{[1]}$ with a space of vector fields $Vect(n).$
Consider $s_k$ as a $k$-ary operation on a space of differential operators
$Dif\!f_n.$ So, $s_k(X_1,\ldots,X_k)$ is a skew-symmetric sum of compositions
of $k$ operators $X_{\sigma(1)}\cdots X_{\sigma(k)}$ by all $k!$ permutations.
In general, composition of $k$ operators of orders $d_1,\ldots,d_k$ is a differential
operator of order $d_1+\cdots+d_k.$ Therefore,
$$X_1\in Dif\!f_n^{[d_1]},\ldots,X_k\in Dif\!f_n^{[d_k]}\Rightarrow s_k(X_1,\ldots,X_k)\in Dif\!f_n^{[d_1+\cdots+d_k]}.$$
In fact differential order of $s_k(X_1,\ldots,X_k)$ is less than $d_1+\ldots+d_k.$ For example, differential order of $s_k(X_1,\ldots,X_k)$ is no more than $n,$ if all $X_1,\ldots,X_k$ are operators of order 1 (vector fields) on $n$-dimensional manifold for any $k$ \cite{odd}.
Moreover, for some $k$ might happen that $s_k$ will be well-defined operation on $Dif\!f_n^{[1]}$:
$$X_1,\ldots,X_k\in Dif\!f_n^{[1]}\Rightarrow
s_k(X_1,\ldots,X_k)\in Dif\!f_n^{[1]}.$$
In \cite{N-commutators} is established  that $s_{n^2+2n-2}$ is well-defined on $Vect(n)=Dif\!f_n^{[1]}$
and in \cite{odd} is proved that $s_{n^2+2n-1}=0$ is
identity on $Vect(n).$
For example, $Vect(2)$ has $6$-commutator and skew-symmetric
identity of degree 7. Space of Hamiltonian vector fields on $2$-dimensional plane
$Vect_0(2)$ has $5$-commutator and skew-symmetric identity of degree 6.

{\bf Question 1.} $(n>1).$ Is it true that $N=n^2+2n-1$ is index of
nilpotency for operator $D,$ i.e., $D^{n^2+2n-1}=0,$ but
$D^{n^2+2n-2}\ne 0$ ?

We think that coefficient at $\prod_{i}\eta_i\prod_{(i,j)\ne
(n,n)}\der_i\eta_j \prod_{i\ne n}\der_i^2\eta_i\der_n$ of
$D^{n^2+2n-2}$ is non-zero. Computer calculations on Mathematica
shows that this coefficient is equal $1, 2,\, 3600$ for $n=2,3,4.$

{\bf Question 2.} $(n>3).$
Is it true that $s_{n^2+2n-2}$ is a unique $N$-commutator
well-defined on $Vect(n),$ for $N>2$?  In other words, is it true that
$$D^N\in Der\,\LL_n, n>3, \Rightarrow N=2 \mbox { or } n^2+2n-2 ?$$

In our paper we prove that for $n=3$ answer to this question is negative.
According to our results  $Vect(3)$ has $2$-commutator, $10$-commutator and $13$-commutator
and this list of $N$-commutators is complete.
Notice that $13$-commutator is connected with a skew-symmetric identity of degree $14,$
but $10$-commutator has no such connection with skew-symmetric
identity of degree 11:
$s_{11}$ even is not well-defined operation on $Vect(3).$

Let us give some quantitive parameters about $D^{10}$ and $D^{13}.$
$D^{10}$ has three escort invariants. They have types $(2,7,1),$ $(3,5,2)$ and $(3,6,0,1)$.
It has $489$ terms of type $(2,7,1),$ $3093$ terms of type $(3,5,2)$,
480 terms of type $(3,6,0,1)$ and all together $4062$ terms.
$D^{13}$ has one escort invariant. It has type $(3,8,2)$ and has $261$ terms.
In other words, $s_{10}(X_1,\ldots,X_{10})$ can be presented as a sum of
$4062$ $10\times 10$-determinants of three types. Similarly, $s_{13}(X_1,\ldots,X_{13})$ can
be presented as a sum of $261$ matrices of order $13\times 13.$

To see that $D^{10}$ is  well-defined on $Vect_0(3),$ we change all terms of $D^{10}$
like $\der^{\alpha}\eta_3\der_3,$ $\alpha_3>0,$
to $-\der^{\alpha-\eps_3+\eps_1}\eta_1\der_1-\der^{\alpha-\eps_3+\eps_2}\eta_2\der_2.$
We obtain element with $864$ terms, among them $82$ 
has type $(2,7,1),$  $76$ has type $(3,6,0,1)$ and $706$ has type
$(3,5,2).$

It is easy to see that $D^k$ is a sum of compositions of the form
$D\star_1 (D\star_2\cdots (D\star_{k-1}D)\cdots),$ where
$(\star_1,\ldots,\star_{k-1})$
is a sequence of two symbols $\circ$ or $\bullet$ such that there are no
two consequative $\bullet$ and whole number of $\bullet$ is no more than
$|I|.$  In particular we see that the differential
order of $D^k$ is no more than $min(k+1/2,|I|).$
This estimate is not strong.
One can see that,  for $n=2,3$ differential orders of $D^k$
are given as follows
$$n=2$$
$$
\begin{array}{cccccccc}
k&1&2&3&4&5&6&7\\
\der deg\, D^k&1&1&2&2&2&1&-\infty\\
\end{array}
$$

$$n=3$$
$$
\begin{array}{ccccccccccccccc}
k& 1&2&3&4&5&6&7&8&9&10&11&12&13&14\\
\der deg\, D^k&1&1&2&2&3&3&3&3&3&1&2&2&1&-\infty\\
\end{array}
$$
\medskip

\noindent If $D\in Der_0\LL_n,$ i.e., $Div\,D=0,$ then the growth of differential orders  of $D^k$ given as

\medskip

$$n=2$$
$$
\begin{array}{cccccccc}
k&1&2&3&4&5&6&7\\
\der deg\, D^k&1&1&2&2&1&-\infty&-\infty\\
\end{array}
$$

$$n=3$$
$$
\begin{array}{ccccccccccccccc}
k& 1&2&3&4&5&6&7&8&9&10&11&12&13&14\\
\der deg\, D^k&1&1&2&2&3&3&3&3&3&1&2&-\infty&-\infty&-\infty\\
\end{array}
$$
We pay attention to a drammatical jumping of
$\der deg\, D^k$ in $(n,k)=(3,10).$
Here we see that $D^{10}$ is a derivation or that the $10$-commutator
is defined correctly on $Vect(3).$
One can check that $Div\,D^{10}=0$ and hence $10$-commutator
is a well defined commutator on $Vect_0(3)$ also.

\begin{th} \label{main1}
Let $D=\sum_{i=1}^3{\eta_i}\der_i\in Der\,\LL_3$  be odd derivation. Then
$$D^{10}\in Der\,\LL_3\otimes {\bf C}[x_1,x_2,x_3],$$
$$D^{13}\in Der\,\LL_3\otimes {\bf C}[x_1,x_2,x_3],$$
$$D^{14}=0.$$
If $D^{N}\in Der\,\LL_3\otimes {\bf C}[x_1,x_2,x_3],$ then $N=2,10,13.$
\end{th}

\begin{th} \label{main2}
Let $D=\sum_{i=1}^3{\eta_i}\der_i\in Der\,\LL_3$  be odd derivation and $Div\,D=\sum_{i=1}^3\der_i\eta_i=0.$  Then
$$D^{10}\in Der\,\LL_3\otimes {\bf C}[x_1,x_2,x_3],$$
$$Div\,D^{10}=0,$$
$$D^{11}=0.$$
If $D^{N}\in Der\,\LL_3\otimes {\bf C}[x_1,x_2,x_3],$ and
$Div\,D=0,$ then $N=2$ or $10.$
\end{th}

\section{ $sl_n$-module structure on $U={\bf C}[x_1,\ldots,x_n]$ and $Dif\!f_n(U)$}

Endow  $U$ by a structure of module over
Lie algebra  $gl_n=<x_i\der_j: i,j=1,\ldots,n, i\ne j>.$
Define an action of $gl_n$ on generators of $U$ by
$$x_i\der_j(\der^{\alpha}(u_s))=
-\delta_{i,s}\der^{\alpha}(u_j)+
\sum_{i=1}^n\alpha_j\der^{\alpha-\epsilon_j+\epsilon_i}(u_s)$$
and prolong this action to $U$ as an even derivation:
$$a(XY)=a(X)Y+Xa(Y),$$
for any $X,Y\in U.$ Prolong the $gl_n$-module structure by natural way
to $Dif\!f_n(U).$ Notice that $gl_n$ acts on $Dif\!f_n(U)$ as a derivation
$$a(F G)=a(F)G+F a(G),$$
and as $gl_n$-module subspaces $<u_i\der_j: i,j=1,\ldots,n>\subset
Dif\!f_n(U)$ and $<\der_i(u_j): i,j=1,\ldots,n>$ are isomorphic to adjoint
module.

Denote by $\pi_1,\ldots,\pi_{n-1}$ fundamental weights of $sl_n$
and by $R(\gamma)$ the irreducible $sl_n$-module with highest weight
$\gamma.$ Let
$${\cal D}^{[s]}=<\der^{\alpha}: \alpha\in {\bf Z}^n_+, |\alpha|=s>$$ and
$$U_s=<\der^{\alpha}u_i: \alpha\in{\bf Z}_+^n,
|\alpha|=s, i=1,\ldots,n>.$$
Since $\der_i$ are even and $u_i$ are odd elements,
take place the following isomorphisms of $sl_n$-modules
$${\cal D}^{[s]}\cong R(s\pi_{1}),$$
$$U_s\cong R(s\pi_1)\cong R(\pi_{n-1}).$$
In particular,
$${\cal D}:=<\der_i : i=1,\ldots,n>\cong R(\pi_{n-1}),$$
$${\cal D}^{[2]}:=<\der_i\der_j: i,j=1,\ldots,n>\cong R(2\pi_1),$$
$$U_0=<u_i: i=1,\ldots,n>\cong R(\pi_1),$$
$$U_{1}=<\der_i(u_j) : i,j=1,\ldots,n>\cong R(\pi_1)\otimes R(\pi_{n-1})
R(\pi_1+\pi_{n-1})\oplus R(0),$$
$$U_{2}=<\der_i\der_j(u_s): i,j,s=1,\ldots,n>\cong
R(2\pi_1)\otimes R(\pi_{n-1})\cong R(2\pi_1+\pi_{n-1})\oplus R(\pi_1).$$
We use the following well-known isomorphisms without special mentioning:
$$\wedge^{n-1}R(\pi_1)\cong R(\pi_{n-1}),$$
$$\wedge^{n}R(\pi_1)\cong R(0),$$
$$\wedge^{n^2-1}R(\pi_1+\pi_{n-1})\cong R(\pi_1+\pi_{n-1}),$$
$$\wedge^{n^2}R(\pi_1+\pi_{n-1})\cong R(0).$$

\begin{lm}\label{askar1}
$a(D^k)=0$ for any $a\in gl_n.$
\end{lm}

{\bf Proof.} If $k=1$ then action of $a\in gl_n$ corresponds
to adjoint derivation and $D$ corresponds to Euler operator.
Therefore,
$$a(D)=[a, \sum_{i=1}^nu_i\der_i]=0.$$
If our statement is true for $k-1$ then
$$a(D^k)=kD^{k-1}[a,D]=0.$$
$\square$

\section{Escort invariants of  $N$-commutators}

Let $L=W_n$ be Witt algebra and $U={\bf C}[x_1,\ldots,x_n]$ be
natural $L$-module. Then
\begin{itemize}
\item
$L=\oplus_{i\ge -1}L_i$ is a graded Lie algebra,
$$L_s=<x^{\alpha}\der_j: |\alpha|=s+1>,$$
\item
$U=\oplus_{i\ge 0} U_i$ be associative commutative graded algebra with $1,$
$$U_s=<x^{\alpha}: |\alpha|=s>,$$
\item
$L$ acts on $U$ as a derivation algebra, i.e.,
$$X(uv)=X(u)v+u(Xv),$$
for any $X\in L, u,v\in U$ and
\item this action is graded:
$$L_iU_j\subseteq U_{i+j}, \quad i\ge -1, j\ge 0.$$
\end{itemize}
In particular, $L_0$ is a Lie algebra isomorphic to $gl_n$ and
all homogeneous components $L_s$ and $U_s$ have structures of $gl_n$-modules.
Then
as $sl_n$-modules,
$$L_{-1}=<\der_i: i=1,\ldots,\der_n>\cong R(\pi_{n-1}),$$
$$L_{0}=<x_i\der_j: i,j=1,\ldots,n>\cong R(\pi_1)\oplus R(\pi_{n-1})\oplus R(0),$$
$$L_1=<x_ix_j\der_s: i,j,s=1,\ldots,n>\cong R(2\pi_1+\pi_{n-1})\oplus R(\pi_1).
$$

Let $M$ be graded $L$-module.
It is called {\it $(L,U)$-module}
if it has additional structure of graded module over
$U$ such that $$X(um)=X(u)m+u X(m),$$
for any $X\in L, u\in U,m\in M.$
Call $M$ $(L,U)$-module with {\it a base $N$} if
$N=M^{L_{-1}}=<m\in M: X(m)=0, \forall X\in L_{-1}>$ and $M$ is free
$U$-module with base $N.$  If $M_1,\ldots, M_k$ and $M$ are
$(L,U)$-modules with bases and $N$ is a base of $M,$ then a space of
polylinear maps $C(M_1,\ldots,M_k;M)=<\psi: M_1\times \cdots \times M_k\rightarrow M>$ is $(L,U)$-module with base and this base as a vector space is
isomorphic to $C(M_1,\ldots,M_k;M).$ In particular, to any
$L_{-1}$-invariant polylinear map $\psi\in C(M_1,\ldots,M_k;M)$ one can
correspond some polylinear map
$esc(\psi)\in C(M_1,\ldots,M_k;N)$ called {\it escort} of $\psi,$
by
$$esc(\psi)(m_1,\ldots,m_k)=pr(\psi(m_1,\ldots,m_k)),$$
where
$$pr: M\rightarrow N,$$
is a projection map to $N,$ i.e., $pr(x^{\alpha} m)=\delta_{\alpha,0}m.$
Inversly, for any $\phi\in C(M_1,\ldots,M_k;N)$ one can correspond some
$L_{-1}$-invariant
polylinear map $\psi=E\phi\in C(M_1,\ldots,M_k;M)$ by
$$E\phi(X_1,\ldots,X_k)=\sum_{a_1\in M_1,\ldots,a_k\in M_k}
E_{a_1}(X_1)\cdots E_{a_k}(X_k)\phi(a_1,\ldots,a_k),$$
where $a_i$ run basic elements of $M_i$ of the form $x^{\alpha} n_i,$
$n_i$ run basic elements of a base of $M_i.$  If $M_1=\cdots =M_k=L$ all
are adjoint modules then
$$E_{x^{\alpha}\der_i}(v\der_j)=\delta_{i,j}\frac{\der^{\alpha}(v)}{\alpha!}.$$
Details of such constructions see \cite{Dzh-Vestnik}.

Apply this theory for $L$-module of differential operators
$M=Dif\!f_n=<u\der^{\alpha}: u\in U, \alpha\in {\bf Z}^n_+>.$
Endow $M=Dif\!f_n$ by grading:
$$M_s=<x^{\alpha}\der^{\beta}: |\alpha|=s>.$$
$Dif\!f_n$ has a structure of associative algebra, in particular, it is a Lie algebra under commutator. As a Lie algebra it has a subalgebra isomorphic to $W_n,$ and hence it has a structure of adjoint module over $W_n.$ Make $Dif\!f_n$ $U$-module under action
$u (v\der^{\beta})=u v\der^{\beta}.$
We see that $M^{L_{-1}}=<\der^{\alpha}: \alpha\in {\bf Z}^n_+>$ and
$M$ is free $U$-module with base $M^{L_{-1}}.$ Therefore, $Dif\!f_n$ is $(L,U)$-module.

Define $s_k\in C^k(L,M)$  by
$$s_k(X_1,\ldots,X_k)=\sum_{\sigma\in Sym_k}sign\,\sigma \,X_{\sigma(1)}\cdots X_{\sigma(k)}.$$
We see that $s_k$ is $L_{-1}$-invariant and graded:
$$\der_i(s_k(X_1,\ldots,X_k))=\sum_{s=1}^ks_k(X_1,\ldots,X_{s-1},[\der_i,X_s],X_{s+1},\ldots,X_k),$$
for any $\der_i\in L_{-1}, X_1,\ldots,X_k\in L,$ and
$$s_k(L_{i_1},\ldots,L_{i_k})\subseteq M_{i_1+\cdots+i_k},$$

Fix some ordering on the set of
basic elements of $W_n.$ Let us take,
for example, the following ordering:
$x^{\alpha}\der_i< X^{\beta}\der_j,$ if $i<j$ or
$|\alpha|<|\beta|$ if
$i=j$ or $\alpha<\beta$ in lexicographic order if $i=j$ and $|\alpha|=|beta|.$
As we mentioned above any $L_{-1}$-invariant cochain $C^k(W_n,Dif\!f_n)$
can be restored by its escort.
In particular, $s_k$ can be restored by its escort.
Any escort is defined as a polylinear map on its support.
Call a subspace of $k$-chains $a_1\wedge \cdots \wedge a_k\in \wedge^k L$
generated by basic vectors  $a_1,\ldots,a_k$ such that
$$s_k(a_1,\ldots,a_k)\in <\der^{\alpha} : \alpha\in {\bf Z}^n_+>$$
as a {\it support} of $s_k.$
Then $supp(s_k)$ has a structure of $sl_n$-module as a
$sl_n$-submodule of $\wedge^kL.$ We know that $sl_n$-module
 $Dif\!f_n^{L_{-1}}=<\der^{\alpha}:
\alpha\in {\bf Z}^n_+>$ is isomorphic to a direct sum of $sl_n$-modules
$R(p\pi_{n-1})$:
$$<\der^{\alpha}: |\alpha|=p, \alpha\in {\bf Z}^n_+>\cong R(p\pi_{n-1})>.$$
Then $supp(s_k)$ is also a direct sum of $sl_n$-submodules $supp_p(s_k),$
where $supp_p(s_k)$ is a $sl_n$-submodule of $\wedge^kL$ generated by
suppport $k$-chains $a_1\wedge\cdots\wedge a_k$ such that
$$s_k(a_1,\ldots,a_k)\in <\der^{\alpha}: \alpha\in {\bf Z}^n_+, |\alpha|=p>.$$

So, we see that any standard skew-symmetric polynomial
$s_k$ induces a serie of $sl_n$-invariant maps
$$supp_p(s_k)\rightarrow R(p\pi_{n-1}).$$
Call such maps {\it escort invariants}.
So, the calculation problem of $k$-commutators is equivalent to the problem of
finding escort invariants.

{\bf Example.}  $esc(s_k)=0$ if $k\ge n^2+2n-1$ and
$s_{n^2+2n-2}$ has exactly one escort invariant $R(\pi_1)\otimes
R(\pi_{n-1}) \otimes \wedge^{n-1}R(2\pi_1+\pi_{n-1})\rightarrow R(\pi_{n-1}).$

\section{Differential polynomials super-agebra $\LL_n$}

Let
${\bf Z}$ be set of integers, ${\bf Z}_+$ a set of non-negative integers,
${\bf Z}^n$ a set of $n$-typles
$\alpha=(\alpha_1,\ldots,\alpha_n), \alpha_i\in {\bf Z}, i\in I,$ and ${\bf Z}^n_+=\{\alpha\in {\bf Z}^n |\alpha_i\ge 0, i\in I\}.$
Let $\eps_i\in {\bf Z}^n$ with $i$-th component 1 and other components are 0.
Then $\alpha=\sum_{i=1}^n\alpha_i\eps_i,$ for any $\alpha\in {\bf Z}^n.$
Set
$$|\alpha|=\sum_{i=1}^n\alpha_i.$$
Endow sets ${\bf Z}^n_+$ and ${\bf Z}^n_+\times \{1,\ldots,n\}$
by linear ordering:
$\alpha<\beta$, if
$$|\alpha|<|\beta|$$ or
$$|\alpha|=|\beta|, \alpha_1=\beta_1,\ldots, \alpha_{s-1}=\beta_{s-1}, \alpha_s>\beta_s,$$
for some $s=1,\cdots,n.$
Set $(\alpha,i)<(\beta,j),$ if $i<j$ or $i=j, \alpha<\beta.$

Let $\LL_{n}$ be an super-commutative associative algebra over a field $K$
generated by odd elements $e_{\alpha,i},$ where
$\alpha\in {\bf Z}^n_+, i\in I.$ Then
$$e_{\alpha,i}e_{\beta,j}=-e_{\beta,j}e_{\alpha_,i},$$
$$e_{\alpha,i}(e_{\beta,j}e_{\gamma,s})=(e_{\alpha,i}e_{\beta,j})e_{\gamma,s},$$
for any $\alpha,\beta,\gamma\in {\bf Z}^n_+, i,j,s\in I.$ Elements
$e_{\alpha,i}e_{\beta,j}\cdots e_{\gamma,s}$ with $(\alpha,i)<(\beta,j)<\cdots<(\gamma,s)$ form  base of $\LL_n.$ We fix this base and call such elements {\it base} elements of $\LL_n$. Call number of indexes $i,j,\cdots,s$ of base element $e$  as its {\it length} and denote $l(e).$

Any base element of $\LL_n$ can be presented as  $e=e^{[-1]} e^{[0]} e^{[1]}\cdots e^{[r]},$ where $e^{[s]}$ is a product of ordered generators of a form $e_{\alpha,i}$ with
$|\alpha| =s+1.$ Call $e^{[s]}$ {\it $s$-component} of $e$ and
its length $l(e^{[s]})$, denote it $l_s(e)$, call  as  {\it $s$-Length} of $e.$
Thus,
$$l(e)=\sum_{i\ge -1}l_{i}(e).$$

Let $\der_i=\frac{\der}{\der_i}, i\in I,$ are partial derivations of
$U=K[x_1,\ldots,x_n].$ Prolong these maps to maps of  $\LL_n$ by
$$\der_ie_{\beta,j}=e_{\alpha+\eps_i,j}.$$
It is easy to see that
$\der_i$ satisfies Leibniz rule
$$\der_i(e_{\beta,j}e_{\gamma,s})=(\der_ie_{\beta,j}) e_{\gamma,s}+e_{\beta,j}(\der_ie_{\gamma,s}),$$
for any $\beta,\gamma\in {\bf Z}^n_+.$
So, we have constructed commuting even derivations
$\der_1,\ldots,\der_n\in Der(\LL_n\otimes U)$
and
$$e_{\alpha,i}=\der^{\alpha} e_{0,i},$$
for any $\alpha\in {\bf Z}^n_+, i\in I.$ Here $0=(0,\ldots,0)\in {\bf Z}^n_+.$

Space $\LL_n$ has three kinds  of gradings.
The first one, ${\bf Z}^n$-grading is defined by
$$||e_{\alpha,i}||= \alpha-\eps_i$$
and for other base elements are prolonged by multiplicativity,
$$||e_{\alpha,i}e_{\beta,j}\cdots e_{\gamma,s}||
=\alpha-\eps_i+\beta-\eps_j+\cdots +\gamma-\eps_s.$$

The second grading is induced by ${\bf Z}^n$-grading.
It is ${\bf Z}$-grading defined on base element $e=e_{\alpha,i}e_{\beta,j}\cdots e_{\gamma,s}$ by
$$|e|=-l(e)+|\alpha|+|\beta|+\cdots +|\gamma|.$$

The third grading is defined by length.
Let $l(\xi)=s,$ if $\xi$ is a nontrivial linear combination of homogeneous base elements of length $s.$

Call
$$wt(e)=|\alpha|+\cdots +|\beta|-l(e)$$ {\it
weight} of $e.$ A parity on $\LL_n$ is defined by length.
Let $\LL_n^{[l]}$ be linear span of base elements $u$ with $l(u)=l.$
Let $\LL_n^{[l,w]}$ be a linear span of base elements $u$ with
$l(e)=l, wt(e)=w.$

{\bf Example.}
$$\LL_{n}^{[1]}=\left< e_{\alpha,i} |\alpha\in
{\bf Z}^n, i=1,\ldots,n\right>,$$ $$
\LL_n^{[n]}=\left<e_{0,1}\cdots e_{0,n}\right>,
$$
$$\LL_n^{[1,-1]}=\left<e_{0,i}\right>.$$

\begin{prp} \label{LL}
$\LL_n$ is associative, super-commutative graded algebra:
$$(uv)w=u(vw),$$
$$uv=(-1)^{q(u)q(v)}vu,$$
$$\LL_n=\oplus_{l\ge 1, w\ge -n }\LL_n^{[l,w]},$$
$$\LL_n^{[l,w]}\LL_n^{l_1,w_1]}\subseteq \LL_n^{[l+l_1,w+w_1]}.$$
for any $u,v,w\in \LL_n.$
\end{prp}

Note that any base element $u\in\LL_n$ can be presented in a form
$u_{-1}u_0\cdots u_{r}$ where $u_s, s=-1,0,\ldots,r$ are base elemenents and
$u_s$ are products of generators
of weight $s.$
We say that base element $u\in \LL_n$ has {\it type} $(l_{-1},l_0,\ldots,l_r)$, if $u$ is a product of $l_s$ generators of weight $s$, for $s=-1,0,\ldots,r.$

\begin{lm}\label{baseLL}
Any base element $u\in \LL_n$ satisfy the following conditions
$$\sum_{i\ge -1}l_{i}(u)=l(u),$$
$$\sum_{i\ge -1} i\,l_{i}(u)=wt(u),$$
$$l_{i}(u)\le n{n+i\choose i+1}, \qquad i\ge -1.$$
\end{lm}

{\bf Proof.} First two relations are reformulations of grading
property of $\LL_n$ (proposition \ref{LL}).
As far as last two relations, they follow from the fact
$$|\{\alpha\in {\bf Z}^n_+ | |\alpha|=i+1\}|={n+i\choose i+1}.$$

{\bf Example.} Let $u=\eta_1\der_1^2\eta_2\der_1\der_2\eta_2.$
Then $u$ is odd base element of type $(1,0,2)$ and $l(u)=3,wt(u)=1.$

Let $Dif\!f_n$ be an algebra of differential operators on $\LL_n$.
It has a base consisting  differential operators of a form $u\der^{\alpha},$ where $\alpha\in {\bf Z}^n_+$ and $u$ is a base element of $\LL_n.$  Endow $Dif\!f_n$ by multiplication $\cdot$ given by
$$u\der^{\alpha}\cdot v\der^{\beta}=\sum_{\gamma}{\alpha\choose\gamma}u\der^{\gamma}v\der^{\alpha+\beta-\gamma}.$$
Here
$${\beta\choose\gamma}=\prod_{i=1}^n{\beta_i+\gamma_i\choose \gamma_i}.$$
Multiplication $\cdot$ corresponds to composition of differential operators.

Endow $Dif\!f_n$ also by two more multiplications $\circ$ and $\bullet.$
They are given by the following rules
$$u\der^{\alpha}\circ v\der^{\beta}=\sum_{\gamma\ne 0}{\alpha\choose\gamma}u\der^{\gamma}v\der^{\alpha+\beta-\gamma},$$
$$u\der^{\alpha}\bullet v\der^{\beta}=u v \der^{\alpha+\beta}.$$
We see that
$$X\cdot Y=X\circ Y+X\bullet Y,$$
for any $X,Y\in Dif\!f_n.$

For a base element $X=u\der^{\alpha}\in Dif\!f_n$ define {\it length}
$l(X)$, {\it weight} $wt(X)$, {\it parity} $q(X)$ and {\it differential
order} $\der deg(X)$ by
$$l(X)=l(u),$$
$$wt(X)=wt(u)+|\alpha|,$$
$$q(X)=l(u),$$
$$\der deg(X)=|\alpha|.$$
Let
$$Dif\!f_n^{[d]}=\left< X| \der deg(X)=d\right>.$$
$$Dif\!f_n^{[l,w]}
=\left<X | l(X)=l, wt(X)=w\right>,$$
$$Dif\!f_n^{[l,w,d]}
=\left<X | l(X)=l, wt(X)=w, \der deg(X)=d\right>.$$
Denote a space of differential operators of first
order $Dif\!f_n^{[1]}$ by $\WW_n.$

For a differential operator  $X=\sum_{\alpha\in {\bf Z}^n_+}v_{\alpha}\der^{\alpha}\in Dif\!f_n,$ define its differential order $deg(X)$
as maximal $|\alpha|$, such that $v_{\alpha}\ne 0.$

\begin{prp} \label{Diff} Space of differential operators under different multiplications have the following properties.

The algebra $(Dif\!f_n,\cdot)$ is associative
super-algebra:
$$X\cdot(Y\cdot Z)=(X\cdot Y)\cdot Z,$$
for any $X,Y,Z\in Dif\!f_n.$ This agebra is graded,
$$Dif\!f_n=\oplus_{l>0, w\ge -n} Dif\!f_n^{[l,w]},$$
$$Dif\!f_n^{[l,w]}\cdot Dif\!f_n^{[l_1,w_1]}\subseteq Dif\!f_n^{[l+l_1,w+w_1]}.$$

The algebra  $(\WW_n,\circ)$ is super-left-symmetric:
$$(X,Y,Z)=(-1)^{q(X)q(Y)}(Y,X,Z),$$
for any differential operators of first order $X,Y,Z$, where
$(X,Y,Z)=X\circ(Y\circ Z)-(X\circ Y)\circ Z$ is associator.
Moreover, super-left-symmetric rule is true  for any $X,Y\in Dif\!f_n^{[1]}, Z\in Dif\!f_n.$ This algebra is graded,
$$\WW_n=\oplus_{l>0, w\ge -n} \WW_n^{[l,w]},$$
$$\WW_n^{[l,w]}\circ \WW_n^{[l_1,w_1]}\subseteq \WW_n^{[l+l_1,w+w_1]}.$$

The algebra $(Dif\!f_n,\bullet)$ is associative super-commutative:
$$X\bullet Y=(-1)^{q(X)q(Y)}Y\bullet X,$$
$$X\bullet(Y\bullet Z)=(X\bullet Y)\bullet Z,$$
for any $X,Y,Z\in Dif\!f_n.$ This algebra is graded under length, weight and
differential order,
$$Dif\!f_n=\oplus_{l>0, w\ge -n,d\ge 0} Dif\!f_n^{[l,w,d]},$$
$$Dif\!f_n^{[l,w,d]}\bullet Dif\!f_n^{[l_1,w_1,d_1]}\subseteq
Dif\!f_n^{[l+l_1,w+w_1,d+d_1]}.$$

Any differential operator of first order under multiplication $\circ$ acts on $(Dif\!f_ n,\bullet)$ as a derivation:
$$X\circ(Y\bullet Z)=(X\circ Y)\bullet Z+(-1)^{q(X)q(Y)}Y\bullet(X\circ Z),$$
for any $X\in \WW_n, Y,Z\in Dif\!f_n.$
\end{prp}

{\bf Proof.}
Notice that natural action of $\WW_n$ on $\LL_n$ coincides
with left-symmetric product:
$$X(\eta)=X\circ \eta,$$
for any $X\in \WW_n, \eta\in \LL_n.$
Therefore, we have the following
connection between composition and left-symmetric multiplications:
$$(X\cdot Y)(\eta)\ne (X\circ Y)(\eta))$$
but
$$(X\cdot Y)(\eta)=X\circ Y(\eta),$$
for any $X,Y\in Dif\!f_n,
\eta\in \LL_n.$
Moreover,composition of differential operators of first order can be
expressed in terms of left-symmetric multiplication,
$$(X\cdot Y)(\eta)=X\circ(Y\circ \eta),$$
for any  $X,Y\in \WW_n, \eta\in \LL_n.$
Thus,
$$(X\circ Y+X\bullet Y)(\eta)=X\circ (Y\circ \eta),$$
and
$$X\circ (Y\circ \eta)-(X\circ Y)(\eta)=(X\bullet Y)(\eta).$$
Since $X\circ Y\in \WW_n,$ this means that
\begin{equation}\label{bir}
X\circ (Y\circ \eta)-(X\circ Y)\circ \eta=(X\bullet Y)(\eta).
\end{equation}
for any  $X,Y\in \WW_n,$ $\eta\in \LL_n.$
By these facts we see that
$$([X,Y]\cdot Z)(\eta)=(X\cdot Y-(-1)^{q(X)q(Y)}Y\cdot X)(Z(\eta))$$
$$=(X\circ Y+X\bullet Y-(-1)^{q(X)q(Y)}Y\circ X-(-1)^{q(X)q(Y)}Y\bullet
X)\circ (Z(\eta))$$
$$=(X\circ Y-(-1)^{q(X)q(Y)}Y\circ X)\circ
(Z(\eta)).$$
On the other hand
$$([X,Y]\cdot Z)(\eta)=(X\cdot (Y\cdot Z)-(-1)^{q(X)q(Y)}
Y\cdot(X\cdot Z))(\eta)$$
$$=X\circ (Y\cdot Z)(\eta)-(-1)^{q(X)q(Y)}Y\circ (X\cdot Z)(\eta)$$
$$=X\circ (Y\circ Z(\eta))-(-1)^{q(X)q(Y)}Y\circ(X\circ Z(\eta)).$$
Hence,
$$(X\circ Y-(-1)^{q(X)q(Y)}Y\circ X)\circ
(Z(\eta))=
X\circ (Y\circ Z(\eta))-(-1)^{q(X)q(Y)}Y\circ(X\circ Z(\eta)).$$
In other words,
$$(X\circ Y-(-1)^{q(X)q(Y)}Y\circ X)\circ Z=
X\circ (Y\circ Z)-(-1)^{q(X)q(Y)}Y\circ(X\circ Z),$$
for any $X,Y\in\WW_n, Z\in Dif\!f_n.$

Other statements of our proposition are evident. $\square$

For a base element $X=u\der^{\alpha}\in Dif\!f_n$ say that it has
{\it type} $(l_{-1},l_0,l_1,\ldots,l_r;d)$ if $u$ has type $(l_{-1},l_0,\ldots,l_r)$ and $|\alpha|=d.$

{\bf Example.} Let $X=\eta_1\eta_3 \der_1\eta_1\der_2\eta_1\der_2\eta_2
\der_1\der_2\der_3 \eta_3\der_1\der_2.$ Then $X$ is base element of $Dif\!f_3$
of type $(2,3,0,1;1),$ weight $2$ and differential order $2.$

\begin{lm}\label{baseDiff}
Any base element $X\in Dif\!f_n$ satisfies the following conditions:
$$\sum_{i\ge -1}l_{i}(X)=l(X),$$
$$\sum_{i\ge -1} i\,l_{i}(X)+deg(X)=wt(X),$$
$$l_{i}(X)\le n{n+i\choose i+1}, \qquad i\ge -1.$$
\end{lm}

{\bf Proof.} Follows from proposition \ref{Diff} and Lemma \ref{baseLL}.
$\square$

Let $Dif\!f_{n}^{(l_{-1},l_0,\ldots,l_r;d)}$ be a subspace of $Dif\!f_n$
generated by base elements of type $(l_{-1},l_0,\ldots,l_r;d).$
Let
$$\tau_{(l_{-1},l_0,\ldots,l_r;d)}: Dif\!f_n\rightarrow Dif\!f_n^{(l_{-1},l_0,\ldots,l_r;d)},$$
$$\tau_d: Dif\!f_n\rightarrow Dif\!f_n^{[d]}$$
be projection maps.

 Polynomial space $U=K[x_1,\ldots,x_n]$ has natural gradings:
$$||x^{\alpha}||=\alpha,\qquad |x^{\alpha}|=|\alpha|.$$
It has standard base $\{x^{\alpha}=\prod_{i=1}^nx_i^{\alpha_i} | \alpha\in {\bf Z}^n\}.$ These gradings on $\LL_n$ and $U$ induce gradings on
$\LL_n\otimes U.$

In previous section we define parity $q$ on $\LL_n\otimes U$.
Below we set
$$\eta_i=e_{0,i}.$$
So, instead of $e_{\alpha,i}$ we can  write $\der^{\alpha}\eta_i.$
Then for
$\eta=\der^{\alpha_1}\eta_{i_1}\cdots\der^{\alpha_k}\eta_{i_k}$
we have
$$l(\eta)=k.$$ We identitfy $\LL_n$ with
$\LL_n\otimes 1$ and consider $\LL_n$ as a subalgebra of $\LL_n\otimes U.$

\section{Differential operators of first order on $\LL_n$}

$\WW_n=Dif\!f_n^{[1]}$ has two algebraic structures.
The first one, a structure of super-Lie algebra, is well-known. Let
$$[D_1,D_2]=D_1D_2-(-1)^{q(D_1)q(D_2)}D_2D_1.$$
be super-commutator.
Then
$$[D_1,D_2]=-(-1)^{q(D_1)q(D_2)}[D_2,D_1],$$
$$[D_1,[D_2,D_3]]=[[D_1,D_2],D_3]+(-1)^{
q(D_1)q(D_2)}[D_2,[D_1,D_3]].$$
Notice that
$$q(\xi\der_i)=q(\xi),$$
for any $\xi\in \LL_n.$
Recall that for any $D\in \WW_n,$ corresponding adjopint operator $$ad\,D:
\WW_n\Rightarrow \WW_n$$ is a derivation of $\WW_n.$ Therefore, $\WW_n$ can
be interpretered as a derivation super-Lie algebra of $\LL_n.$

The second structure of algebra on $\WW_n$ can be done by left-symmetric
multiplication. It is less known.  Define a product $\circ $ by
$$(\xi\der_i)\circ (\eta\der_j)=\xi\der_i(\eta)\der_j.$$
Then for ant $D_1,D_2,D_3\in \WW_n,$
$$(D_1,D_2,D_3)=(-)^{q(D_1)q(D_2)}(D_2,D_1,D_3)$$
(left-symmetric identity). Here
$$(D_1,D_2,D_3)=D_1\circ(D_2\circ D_3)-(D_1\circ D_2)\circ D_3$$
is associator.

{\bf Remark.} Let $Dif\!f_n^{(k)}$ be subspace of $Dif\!f_n$ of order
no more than $k.$ Well known that
$$Dif\!f_0^{(0)}=U\subset Dif\!f_n^{(1)}\subset Dif\!f_n^{(2)}\subset \cdots$$
is an increasing filtration on $Dif\!f_n,$
$$Dif\!f_n^{(k)}\cdot Dif\!f_n^{(s)}\subseteq Dif\!f_n^{(k+s)},\qquad  k,s \ge 0.$$
So, $Dif\!f_n^{(k)}$ has a structure of algebra under composition operation,
if $k=0,$ $Dif\!f_n^{(1)}$ has an algebraic structure under commutator.
One cas ask about algebraic structures on $Dif\!f_n^{(k)}$ for $k>0.$
In other words, is it possible to find some $N=N(n,k)$,  such that
$$X_1,\ldots,X_N\in Dif\!f_n^{(k)}\Rightarrow s_N(X_1,\ldots,X_N)\in Dif\!f_n^{(k)}.$$

One can prove the following

\begin{th} Let $n>1.$ Then $s_{(n+1)^2}=0$ is identity on $Dif\!f_n^{(1)}$ and
$s_{n^2+2n}, s_{n^2+2n-1}$ are well-defined operations on $Dif\!f_n^{(1)}.$
Moreover,
$s_{n^2+2n}(X_1,\ldots,X_{n^2+2n})\in Dif\!f_n^{(0)},$ for all $X_1,\ldots,X_{n^2+2n}\in Dif\!f_n^{(1)}.$
\end{th}

\section{Calculation of  $D^n$}

Let $\eta_1,\ldots,\eta_n$ are odd elements and
$$D=\sum_{i=1}^n\eta_i\der_i,$$
$$F=D\circ D=\sum_{i,j=1}^n \eta_i\der_i\eta_j\der_j..$$
Notice that
\begin{itemize}
\item $D\in \WW_n^{[1,0]}$
\item $F$ is even element of $\WW_n$
\item $l_{-1}(F)=1, l_{0}(F)=1, l_s(F)=0, s>0.$
\end{itemize}
Therefore,
$$D^k\in Dif\!f_n^{[k,0]}.$$

Define left-symmetric power  $D^{\circ k}$ by
$$\begin{array}{ll}
D^{\circ k}=&D\circ D^{\circ (k-1)},\mbox{ if } k>1,\\
D^{\circ 1}=&D \qquad .
\end{array}$$
Similarly one defines bullet power $D^{\bullet k}$ and associative power
$D^{\cdot k}.$ Since multiplications $\cdot$ and $\bullet$ are associative,
in last cases $D^{\bullet k}$ and $D^{\cdot k}$ have usual properties of
powers
$$D^{\bullet k}\bullet D^{\bullet s}=D^{\bullet (k+s)},$$
$$D^{\cdot k}\bullet D^{\cdot s}=D^{\cdot (k+s)},$$
These facts are not true for left-symmetric powers. For example,
$$D\circ (D\circ D^{\circ 2})=(D\circ D)\circ D^{\circ 2},$$
but
$$D\circ D^{\circ 2}\circ D)\ne (D\circ D^{\circ 2})\circ D.$$

\begin{lm}\label{D^2=F}
$D^{\cdot 2}=F.$
\end{lm}

{\bf Proof.}
$$D^{\cdot 2}=D\cdot D=\sum_{i,j=1}^n
\eta_i\der_i\eta_j\der_j+\sum_{i,j=1}^n\eta_i\eta_j\der_i\der_j.
$$
Since $\eta_i\eta_j=-\eta_j\eta_i$ and
$\der_i\der_j=\der_j\der_i,$ we have
$$\sum_{i,j=1}^n
\eta_i\eta_j\der_i\der_j=0.$$
Thus,
$$D^2=
\sum_{i,j=1}^n\eta_i\der_i\eta_j\der_j
=D\circ D=F.$$

\begin{lm}\label{D2n}
$D^{\circ (2n)}=F^{\circ n}$ for any $n=1,2,3,\cdots $
\end{lm}

{\bf Proof.} We use induction on $n.$

If $n=1,$ then nothing is to prove.

Suppose that
$$D^{\circ (2(n-1))}=F^{\circ (n-1)}$$
for some $n>1.$ Then by definition
$$D^{\circ (2n)}=D\circ (D\circ D^{\circ 2(n-1)})$$
Since $D$ is odd, by left-symmetric property of $(\WW_n,\circ)$
(proposition \ref{Diff})
$$(D,D,G)=$$
for any $G\in \WW_n.$ Thus,
$$D\circ (D\circ G)=(D\circ D)\circ G.$$
Therefore,
$$D^{\circ (2n)}=(D\circ D)\circ D^{\circ (2(n-1))}.$$
By inductive suggestion,
$$D^{\circ (2n)}= F\circ F^{\circ (n-1)}=F^{\circ n}.$$

\begin{lm}\label{FFF}
$F\circ F^{\bullet k}=k F^{\bullet (k-1)}\bullet F^{\circ 2}$
\end{lm}

{\bf Proof.} Since $F\in \WW_n$ is even derivation,
any any left-symmetric multiplication operator acts on $(Dif\!f_n,\bullet)$
as a super-derivation (proposition \ref{Diff}) we have
$$F\circ (F^\bullet F)=(F\circ F)\bullet F+F\bullet (F\circ F).$$
By commutativity of bullet-multiplication this means that
$$F\circ F^{\bullet 2}=2 F\bullet F^{\circ 2}.$$
Easy induction on $k$ based on a such arguments shows
that our lemma is true in general case.

\begin{lm}\label{D^4}
$D^4=F^{\circ 2}+F^{\bullet 2}$
\end{lm}

{\bf Proof.} By Lemma \ref{D^2=F} and by associativity of $\circ$,
$$D^4=D^2\cdot D^2=F\cdot F=F\circ F+F\bullet F.$$

\begin{lm}\label{D^6}
$D^6= F^{\circ 3}+3 F\bullet F^{\circ 2}+F^{\bullet 3}.$
\end{lm}

{\bf Proof.} By Lemma \ref{D^2=F} and Lemma \ref{D^4},
$$D^6=D^2\cdot D^4=D^2\circ D^4+D^2\bullet D^4$$
$$=F\circ (F^{\circ 2}+F^{\bullet 2})+F\bullet (F^{\circ 2}+F^{\bullet 2})$$
$$F^{\circ 3}+F\circ F^{\bullet 2}+F\bullet F^{\circ 2}+F^{\bullet 3}.$$
Thus by Lemma \ref{FFF},
$$D^6=F^{\circ 3}+3 F\bullet F^{\circ 2}+F^{\bullet 3}.$$

\begin{lm} \label{D^8}
$D^8=F^{\circ 4}+3 F^{\circ 2}\bullet F^{\circ 2}+4 F\bullet F^{\circ 3}
+6 F^{\circ 2}\bullet F^{\bullet 2}+F^{\bullet 4}.$
\end{lm}

{\bf Proof.} By Lemma \ref{FFF}
$$F\circ F^{\bullet 3}=3 F^{\circ 2}\bullet  F^{\bullet 2}.$$
Therefore, by Lemma \ref{D^2=F}, Lemma \ref{D^6}
$$D^8=D^2\cdot D^6=D^2\circ D^6+D^2\bullet D^6$$

$$=F\circ (F^{\circ 3}+3 F\bullet F^{\circ 2}+F^{\bullet 3})$$
$$+F\bullet (F^{\circ 3}+3 F\bullet F^{\circ 2}+F^{\bullet 3})$$

$$=F^{\circ 4}+3 F^{\circ 2}\bullet F^{\circ 2}+3 F\bullet F^{\circ 3}
+3 F^{\circ 2}\bullet F^{\bullet 2}$$
$$ +F\bullet F^{\circ 3}+3F^{\bullet 2}\bullet F^{\circ 2}+F^{\bullet 4}$$

$$=F^{\circ 4}+3 F^{\circ 2}\bullet F^{\circ 2}+4 F\bullet F^{\circ 3}
+6 F^{\circ 2}\bullet F^{\bullet 2}+F^{\bullet 4}.$$

\begin{lm}\label{D^{10}}
$$D^{10}=
F^{\circ 5}$$
$$+5(F^{\circ 2}\bullet F^{\circ 3}+ F\circ (F\bullet F^{\circ 3}))$$
$$+5(2 F^{\circ 3}\bullet F^{\bullet 2}
+3 F\bullet F^{\circ 2}\bullet F^{\circ 2})
$$ $$+4 F^{\circ 2}\bullet F^{\bullet 3}
+6 F^{\circ 2}\bullet F^{\bullet 3}+F^{\bullet 5}
$$
\end{lm}

{\bf Proof.} By Lemma \ref{D^8} and Lemma \ref{FFF}
$$D^{10}=D^2\circ D^8=D^2\circ D^8+D^2\bullet D^8$$

$$=F\circ (
F^{\circ 4}+3 F^{\circ 2}\bullet F^{\circ 2}+4 F\bullet F^{\circ 3}
+6 F^{\circ 2}\bullet F^{\bullet 2}+F^{\bullet 4})$$
$$+F\bullet (
F^{\circ 4}+3 F^{\circ 2}\bullet F^{\circ 2}+4 F\bullet F^{\circ 3}
+6 F^{\circ 2}\bullet F^{\bullet 2}+F^{\bullet 4})$$

$$=F^{\circ 5}+3 F\circ (F^{\circ 2}\bullet F^{\circ 2})+4 F\circ (F\bullet
F^{\circ 3})+6 F\circ(F^{\circ 2}\bullet F^{\bullet 2})+F\circ (F^{\bullet
4})$$
$$+F\bullet
F^{\circ 4}+
3 F\bullet F^{\circ 2}\bullet F^{\circ 2}+4 F^{\bullet 2}\bullet F^{\circ 3}
+6 F^{\circ 2}\bullet F^{\bullet 3}+F^{\bullet 5})$$

$$=F^{\circ 5}
+6 F^{\circ 2}\bullet F^{\circ 3}+4F^{\circ 2}\bullet F^{\circ 3}+4 F\bullet
F^{\circ 4}$$
$$+6 F^{\circ 3}\bullet F^{\bullet 2}+
12 F^{\circ 2}\bullet F^{\circ 2}\bullet F+4 F^{\circ 2}\bullet F^{\bullet
3}$$
$$+F\bullet
F^{\circ 4}+3 F\bullet F^{\circ 2}\bullet F^{\circ 2}+4 F^{\bullet 2}\bullet
F^{\circ 3}
+6 F^{\circ 2}\bullet F^{\bullet 3}+F^{\bullet 5})$$

$$=F^{\circ 5}$$ $$
+10 F^{\circ 2}\bullet F^{\circ 3}
+5 F\bullet F^{\circ 4}$$ $$
+10 F^{\circ 3}\bullet F^{\bullet 2}
+15 F\bullet F^{\circ 2}\bullet F^{\circ 2}$$ $$
+4 F^{\circ 2}\bullet F^{\bullet 3}
+6 F^{\circ 2}\bullet F^{\bullet 3}$$ $$+F^{\bullet 5}.$$

By Lemma \ref{FFF},
$$F\circ (F^{\circ 3}\bullet F)=F^{\circ 4}\bullet F+F^{\circ 3}
\bullet F^{\circ 2}.$$
Thus
$$2 F^{\circ 2}\bullet F^{\circ 3}+F\bullet F^{\circ 4}=$$
$$F^{\circ 2}\bullet F^{\circ 3} +F\circ (F\bullet F^{\circ 3}).$$
Our lemma is proved.

\begin{lm} \label{Altynai} For any $G\in Dif\!f_n,$
$$F\circ (\prod_{r=1}^n\eta_r \; G)=0.$$
\end{lm}

{\bf Proof.} We have $$\eta_i\der_i(\eta_j)\der_j(\eta_1\cdots \eta_n)$$
$$=\sum_{s=1}^n \xi_s,$$
where $$\xi_s=
\eta_i\der_i\eta_j \eta_1\cdots\eta_{s-1}\der_j(\eta_s)\eta_{s+1}\cdots
\eta_n.
$$ If $s\ne i,$ then
$$ \xi_s=\pm \eta_i\eta_i \xi_{i,s},$$ where
$$\xi_{i,s}=
\der_i\eta_j\der_j\eta_s \prod_{r\ne i,s} \eta_r.$$
Since $\eta_i\eta_i=0,$ this means that
$$\xi_s=0,$$
if $s\ne i.$ If $s=i,$ then
$$\xi_s=\pm \eta_i\der_i\eta_j\der_j\eta_i\prod_{r\ne i}\eta_r=
\der_i\eta_j\der_j\eta_i(\prod_{r}\eta_r).
$$
We have
$$\sum_{i,j=1}^n\der_i\eta_j\der_j\eta_i=\theta_1+\theta_2+\theta_3,$$ where
$$\theta_1=\sum_{i<j}\der_i\eta_j\der_j\eta_i,$$
$$\theta_2=\sum_i\der_i\eta_i\der_i\eta_i,$$
$$\theta_3=\sum_{i>j}\der_i\eta_j\der_j\eta_i,$$
Since elements
$\der_i\eta_j$ and $\der_j\eta_i$ are odd,
$$\theta_1+\theta_3=0, \qquad \theta_2=0.$$
Thus,
$$F\circ (\prod_{r=1}^n \eta_r G)=(\sum_{i,j=1}^n
\der_i\eta_j\der_j\eta_i) \prod_{r}\eta_r G
=0.$$

Let
$$Dif\!f_n^{[s]}=< u\der^{\alpha} | u\in \LL_n, \alpha\in \Gamma_n,
|\alpha|=s>$$
be a space of differential operators of order $s$ and
$$\tau_s: Dif\!f_n\rightarrow Dif\!f_n^{[s]}$$
be projection map.

\begin{lm} \label{mainlemma} If $n=3,$ $D=\sum_{i=1}^n u_i\der_i,$ and $u_i$
are odd, then
$$\tau_1 D=F^{\circ 5},$$
$$\tau_2D^{10}=
5(F^{\circ 2}\bullet F^{\circ 3}+F\circ (F\bullet F^{\circ 3})),$$
$$\tau_3D^{10}=5(2 F^{\circ 3}\bullet F^{\bullet 2}+3 F\bullet F^{\circ
2}\bullet F^{\circ 2}),$$
$$\tau_sD^{10}=0, \qquad s>3.$$
\end{lm}

{\bf Proof.} Follows from Lemma \ref{D^{10}} and from the fact that
$F^{\bullet s}=0,$  if $s>n.$

{\bf Conclusion.} To find $D^{10}$ we need to calculate $F^{\circ s},$ for
$s=1,2,3$ and $F^{\bullet 2}.$

\section{Second bullet-power of $F$}

The following calculations are not difficult.
$$F^{\bullet 2}_{\eta_1\eta_2;\der_1^2}=
-2\eta_1\eta_2\der_1\eta_1\der_2\eta_1,$$

$$F^{\bullet 2}_{\eta_1\eta_3;\der_1^2}=
-\eta_1\eta_32\der_1\eta_1\der_3\eta_1\der_1^2,$$

$$F^{\bullet 2}_{\eta_2\eta_3;\der_1^2}=
-2\eta_2\eta_3\der_2\eta_1\der_3\eta_1\der_1^2,$$

\section{Second left-symmetric power of $F$
\label{secondlsym}}

It is not hard to obtain the following results.
$$F^{\circ 2}_{\eta_1;\der_1}=$$
$$\eta_1(-2 \der_1\eta_1\der_2\eta_1\der_1\eta_2
-2 \der_1\eta_1\der_3\eta_1\der_1\eta_3
+\der_2\eta_1\der_1\eta_2\der_2\eta_2$$ $$-
  \der_2\eta_1\der_3\eta_2\der_1\eta_3+
  \der_3\eta_1\der_1\eta_2\der_2\eta_3+
  \der_3\eta_1\der_1\eta_3\der_3\eta_3)\der_1,$$

$$F^{\circ 2}_{\eta_2;\der_1}=$$
$$\eta_2(-\der_1\eta_1\der_2\eta_1\der_2\eta_2-
  \der_1\eta_1\der_3\eta_1\der_2\eta_3-
  \der_2\eta_1\der_3\eta_1\der_1\eta_3$$ $$-
  \der_2\eta_1\der_3\eta_2\der_2\eta_3+
  \der_3\eta_1\der_2\eta_2\der_2\eta_3+
  \der_3\eta_1\der_2\eta_3\der_3\eta_3)\der_1,
$$

$$F^{\circ 2}_{\eta_3;\der_1}=$$
$$\eta_3(-\der_1\eta_1\der_2\eta_1\der_3\eta_2-
  \der_1\eta_1\der_3\eta_1\der_3\eta_3-
  \der_2\eta_1\der_2\eta_2\der_3\eta_2$$ $$+
  \der_2\eta_1\der_3\eta_1\der_1\eta_2-
  \der_2\eta_1\der_3\eta_2\der_3\eta_3+
  \der_3\eta_1\der_3\eta_2\der_2\eta_3)\der_1,
$$

$$F^{\circ 2}_{\eta_1\eta_2;\der_1}=$$
$$\eta_1\eta_2(-\der_1\eta_1 \der_1\der_2\eta_1 -
      \der_1\eta_2 \der_2^2   \eta_1 -
      \der_1\eta_3 \der_2\der_3\eta_1 $$ $$+
       \der_2\eta_1 \der_1^2   \eta_1 +
     \der_2\eta_2 \der_1\der_2\eta_1 +
     \der_2\eta_3 \der_1\der_3\eta_1)\der_1,$$

$$F^2(\eta_1\eta_3;\der_1)=$$
$$\eta_1\eta_3(-\der_1\eta_1 \der_1\der_3\eta_1 -
      \der_1\eta_2 \der_2\der_3\eta_1 -
      \der_1\eta_3 \der_3^2   \eta_1 $$ $$+
        \der_3\eta_1 \der_1^2   \eta_1 +
        \der_3\eta_2 \der_1\der_2\eta_1 +
      \der_3\eta_3 \der_1\der_3\eta_1)\der_1,$$

$$F^{\circ 2}_{\eta_2\eta_3;\der_1}=$$
$$ \eta_2\eta_3(- \der_2\eta_1 \der_1\der_3\eta_1 -
     \der_2\eta_2 \der_2\der_3\eta_1 -
       \der_2\eta_3 \der_3^2   \eta_1$$ $$ +
       \der_3\eta_1 \der_1\der_2\eta_1 +
       \der_3\eta_2 \der_2^2   \eta_1 +
      \der_3\eta_3 \der_2\der_3\eta_1)\der_1 . $$

\section{Third left-symmetric power of $F$  \label{thirdlsym}}

In this section we give results of some calculations concerning
$F^{\circ 3}=F\circ(F\circ F)$

$$F^{\circ 3}_{\eta_1;\der_1}=$$
$$\eta_1(
2 \der_1\eta_1\der_2\eta_1\der_1\eta_2\der_3\eta_2\der_2\eta_3+2
        \der_1\eta_1\der_2\eta_1\der_2\eta_2\der_3\eta_2\der_1\eta_3-6
        \der_1\eta_1\der_2\eta_1\der_3\eta_1\der_1\eta_2\der_1\eta_3$$ $$
-2
        \der_1\eta_1\der_2\eta_1\der_3\eta_2\der_1\eta_3\der_3\eta_3-2
        \der_1\eta_1\der_3\eta_1\der_1\eta_2\der_2\eta_2\der_2\eta_3-2
        \der_1\eta_1\der_3\eta_1\der_1\eta_2\der_2\eta_3\der_3\eta_3$$
$$+2
        \der_1\eta_1\der_3\eta_1\der_3\eta_2\der_1\eta_3\der_2\eta_3-2
        \der_2\eta_1\der_1\eta_2\der_2\eta_2\der_3\eta_2\der_2\eta_3+
   \der_2\eta_1\der_1\eta_2\der_3\eta_2\der_2\eta_3\der_3\eta_3$$ $$-
   \der_2\eta_1\der_2\eta_2\der_3\eta_2\der_1\eta_3\der_3\eta_3+2
        \der_2\eta_1\der_3\eta_1\der_1\eta_2\der_1\eta_3\der_3\eta_3-2
        \der_2\eta_1\der_3\eta_1\der_1\eta_2\der_2\eta_2\der_1\eta_3$$ $$+
   \der_3\eta_1\der_1\eta_2\der_2\eta_2\der_2\eta_3\der_3\eta_3-
   \der_3\eta_1\der_2\eta_2\der_3\eta_2\der_1\eta_3\der_2\eta_3-2
\der_3\eta_1\der_3\eta_2\der_1\eta_3\der_2\eta_3\der_3\eta_3)\der_1$$

(all together $15$ terms )

\bigskip
\bigskip

$$F^{\circ 3}_{\eta_2;\der_1}=$$
$$\eta_2(2 \der_1\eta_1\der_2\eta_1\der_2\eta_2\der_3\eta_2\der_2\eta_3-2
        \der_1\eta_1\der_2\eta_1\der_3\eta_1\der_1\eta_2\der_2\eta_3-2
        \der_1\eta_1\der_2\eta_1\der_3\eta_1\der_2\eta_2\der_1\eta_3$$ $$-
\der_1\eta_1\der_2\eta_1\der_3\eta_2\der_2\eta_3\der_3\eta_3-
\der_1\eta_1\der_3\eta_1\der_2\eta_2\der_2\eta_3\der_3\eta_3-2
        \der_2\eta_1\der_3\eta_1\der_1\eta_2\der_2\eta_2\der_2\eta_3$$ $$+
\der_2\eta_1\der_3\eta_1\der_2\eta_2\der_1\eta_3\der_3\eta_3+
  \der_2\eta_1\der_3\eta_1\der_3\eta_2\der_1\eta_3\der_2\eta_3)\der_1.$$

(all together $8$ terms)
\bigskip
\bigskip

$$F^{\circ 3}_{\eta_3;\der_1}=$$
$$\eta_3(
\der_1\eta_1\der_2\eta_1\der_2\eta_2\der_3\eta_2\der_3\eta_3-2
        \der_1\eta_1\der_2\eta_1\der_3\eta_1\der_1\eta_2\der_3\eta_3-2
        \der_1\eta_1\der_2\eta_1\der_3\eta_1\der_3\eta_2\der_1\eta_3$$ $$+
  \der_1\eta_1\der_3\eta_1\der_2\eta_2\der_3\eta_2\der_2\eta_3-2
        \der_1\eta_1\der_3\eta_1\der_3\eta_2\der_2\eta_3\der_3\eta_3-
  \der_2\eta_1\der_3\eta_1\der_1\eta_2\der_2\eta_2\der_3\eta_3$$ $$-
  \der_2\eta_1\der_3\eta_1\der_1\eta_2\der_3\eta_2\der_2\eta_3+2
\der_2\eta_1\der_3\eta_1\der_3\eta_2\der_1\eta_3\der_3\eta_3)\der_1.$$

(all together $8$ terms)
\bigskip
\bigskip

$$F^{\circ 3}_{\eta_1\eta_2;\der_1}=$$
$$\eta_1\eta_2( 3 \der_1\eta_1 \der_1\eta_2 \der_2\eta_2
              \der_2^2   \eta_1 +
        3    \der_1\eta_1 \der_1\eta_2 \der_2\eta_3 \der_2\der_3\eta_1 +
        4    \der_1\eta_1 \der_1\eta_3 \der_2\eta_3 \der_3^2   \eta_1$$ $$ -
         \der_1\eta_1 \der_1\eta_3 \der_3\eta_3 \der_2\der_3\eta_1 -
        2    \der_1\eta_1 \der_2\eta_1 \der_1\eta_2 \der_1\der_2\eta_1 +
         \der_1\eta_1 \der_2\eta_1 \der_1\eta_2 \der_2^2   \eta_2 $$ $$-
        5    \der_1\eta_1 \der_2\eta_1 \der_1\eta_3 \der_1\der_3\eta_1 +
         \der_1\eta_1 \der_2\eta_1 \der_1\eta_3 \der_2\der_3\eta_2 -
        2    \der_1\eta_1 \der_2\eta_1 \der_2\eta_2 \der_1\der_2\eta_2$$ $$
+
         \der_1\eta_1 \der_2\eta_1 \der_2\eta_2 \der_1^2   \eta_1 -
        3    \der_1\eta_1 \der_2\eta_1 \der_2\eta_3 \der_1\der_3\eta_2 +
        3    \der_1\eta_1 \der_2\eta_1 \der_3\eta_1 \der_1^2   \eta_3 $$ $$+
         \der_1\eta_1 \der_2\eta_1 \der_3\eta_2 \der_1\der_2\eta_3 -
        4    \der_1\eta_1 \der_2\eta_2 \der_1\eta_3 \der_2\der_3\eta_1 -
        2    \der_1\eta_1 \der_2\eta_2 \der_2\eta_3 \der_1\der_3\eta_1$$ $$
-
        2    \der_1\eta_1 \der_2\eta_3 \der_3\eta_3 \der_1\der_3\eta_1 +
         \der_1\eta_1 \der_3\eta_1 \der_1\eta_2 \der_2^2   \eta_3 +
        3    \der_1\eta_1 \der_3\eta_1 \der_1\eta_3 \der_1\der_2\eta_1 $$
$$+
         \der_1\eta_1 \der_3\eta_1 \der_1\eta_3 \der_2\der_3\eta_3 -
        3    \der_1\eta_1 \der_3\eta_1 \der_2\eta_2 \der_1\der_2\eta_3 +
         \der_1\eta_1 \der_3\eta_1 \der_2\eta_3 \der_1\der_2\eta_2 $$ $$-
        3    \der_1\eta_1 \der_3\eta_1 \der_2\eta_3 \der_1\der_3\eta_3 +
         \der_1\eta_1 \der_3\eta_1 \der_2\eta_3 \der_1^2   \eta_1 +
         \der_1\eta_1 \der_3\eta_1 \der_3\eta_3 \der_1\der_2\eta_3 $$ $$+
         \der_1\eta_1 \der_3\eta_2 \der_1\eta_3 \der_2^2   \eta_1 +
        2    \der_1\eta_1 \der_3\eta_2 \der_2\eta_3 \der_1\der_2\eta_1 -
        3    \der_1\eta_2 \der_2\eta_2 \der_2\eta_3 \der_2\der_3\eta_1 $$
$$-
        3    \der_1\eta_2 \der_2\eta_3 \der_3\eta_3 \der_2\der_3\eta_1 +
        3    \der_1\eta_2 \der_3\eta_2 \der_2\eta_3 \der_2^2   \eta_1 -
        4    \der_1\eta_3 \der_2\eta_3 \der_3\eta_3 \der_3^2   \eta_1 $$ $$+
        2    \der_2\eta_1 \der_1\eta_2 \der_1\eta_3 \der_2\der_3\eta_1 -
        2    \der_2\eta_1 \der_1\eta_2 \der_2\eta_2 \der_1\der_2\eta_1 +
         \der_2\eta_1 \der_1\eta_2 \der_2\eta_2 \der_2^2   \eta_2 $$ $$+
        2    \der_2\eta_1 \der_1\eta_2 \der_2\eta_3 \der_2\der_3\eta_2 -
         \der_2\eta_1 \der_1\eta_2 \der_3\eta_2 \der_2^2   \eta_3 +
        2    \der_2\eta_1 \der_1\eta_3 \der_2\eta_3 \der_3^2   \eta_2$$ $$ +
        3    \der_2\eta_1 \der_1\eta_3 \der_3\eta_3 \der_1\der_3\eta_1 +
         \der_2\eta_1 \der_2\eta_2 \der_1\eta_3 \der_1\der_3\eta_1 -
         \der_2\eta_1 \der_2\eta_2 \der_1\eta_3 \der_2\der_3\eta_2 $$ $$-
         \der_2\eta_1 \der_2\eta_2 \der_2\eta_3 \der_1\der_3\eta_2 +
         \der_2\eta_1 \der_2\eta_2 \der_3\eta_2 \der_1\der_2\eta_3 +
        2    \der_2\eta_1 \der_3\eta_1 \der_1\eta_2 \der_1\der_2\eta_3 $$
$$+
        2    \der_2\eta_1 \der_3\eta_1 \der_1\eta_3 \der_1\der_3\eta_3 -
        4    \der_2\eta_1 \der_3\eta_1 \der_1\eta_3 \der_1^2   \eta_1 -
         \der_2\eta_1 \der_3\eta_1 \der_2\eta_3 \der_1^2   \eta_2 $$ $$-
         \der_2\eta_1 \der_3\eta_1 \der_3\eta_3 \der_1^2   \eta_3 -
        2    \der_2\eta_1 \der_3\eta_2 \der_1\eta_3 \der_1\der_2\eta_1 +
         \der_2\eta_1 \der_3\eta_2 \der_1\eta_3 \der_2\der_3\eta_3 $$ $$-
         \der_2\eta_1 \der_3\eta_2 \der_2\eta_3 \der_1\der_3\eta_3 -
         \der_2\eta_1 \der_3\eta_2 \der_2\eta_3 \der_1^2   \eta_1 +
        2    \der_2\eta_2 \der_1\eta_3 \der_2\eta_3 \der_3^2   \eta_1 $$ $$+
        2    \der_2\eta_2 \der_1\eta_3 \der_3\eta_3 \der_2\der_3\eta_1 +
         \der_2\eta_2 \der_2\eta_3 \der_3\eta_3 \der_1\der_3\eta_1 -
         \der_2\eta_2 \der_3\eta_2 \der_1\eta_3 \der_2^2   \eta_1$$ $$ -
        2    \der_2\eta_2 \der_3\eta_2 \der_2\eta_3 \der_1\der_2\eta_1 -
        2    \der_3\eta_1 \der_1\eta_2 \der_1\eta_3 \der_2^2   \eta_1 +
        2    \der_3\eta_1 \der_1\eta_2 \der_2\eta_2 \der_2^2   \eta_3$$ $$ -
        2    \der_3\eta_1 \der_1\eta_2 \der_2\eta_3 \der_1\der_2\eta_1 +
        2    \der_3\eta_1 \der_1\eta_2 \der_2\eta_3 \der_2\der_3\eta_3 -
         \der_3\eta_1 \der_1\eta_2 \der_2\eta_3 \der_2^2   \eta_2 $$ $$-
         \der_3\eta_1 \der_1\eta_2 \der_3\eta_3 \der_2^2   \eta_3 -
        4    \der_3\eta_1 \der_1\eta_3 \der_2\eta_3 \der_1\der_3\eta_1 -
         \der_3\eta_1 \der_1\eta_3 \der_2\eta_3 \der_2\der_3\eta_2 $$ $$+
        2    \der_3\eta_1 \der_1\eta_3 \der_2\eta_3 \der_3^2   \eta_3 -
         \der_3\eta_1 \der_1\eta_3 \der_3\eta_3 \der_1\der_2\eta_1 -
         \der_3\eta_1 \der_1\eta_3 \der_3\eta_3 \der_2\der_3\eta_3 $$ $$+
        3    \der_3\eta_1 \der_2\eta_2 \der_1\eta_3 \der_1\der_2\eta_1 -
        2    \der_3\eta_1 \der_2\eta_2 \der_1\eta_3 \der_2\der_3\eta_3 +
         \der_3\eta_1 \der_2\eta_2 \der_2\eta_3 \der_1\der_2\eta_2 $$ $$+
         \der_3\eta_1 \der_2\eta_2 \der_2\eta_3 \der_1^2   \eta_1 +
         \der_3\eta_1 \der_2\eta_2 \der_3\eta_3 \der_1\der_2\eta_3 +
         \der_3\eta_1 \der_2\eta_3 \der_3\eta_3 \der_1\der_3\eta_3 $$ $$+
         \der_3\eta_1 \der_2\eta_3 \der_3\eta_3 \der_1^2   \eta_1 -
        3    \der_3\eta_2 \der_1\eta_3 \der_2\eta_3 \der_2\der_3\eta_1 -
         \der_3\eta_2 \der_1\eta_3 \der_3\eta_3 \der_2^2   \eta_1$$ $$ +
         \der_3\eta_2 \der_2\eta_3 \der_3\eta_3 \der_1\der_2\eta_1)\der_1.$$

(all together $76$ terms)

\bigskip
\bigskip

$$F^{\circ 3}_{\eta_1\eta_3;\der_1}=$$
$$\eta_1\eta_3(- \der_1\eta_1 \der_1\eta_2 \der_2\eta_2 \der_2\der_3\eta_1
-
         \der_1\eta_1 \der_1\eta_2 \der_2\eta_3 \der_3^2   \eta_1 +
        4    \der_1\eta_1 \der_1\eta_2 \der_3\eta_2 \der_2^2   \eta_1 $$ $$+
        4    \der_1\eta_1 \der_1\eta_2 \der_3\eta_3 \der_2\der_3\eta_1 +
        3    \der_1\eta_1 \der_1\eta_3 \der_3\eta_3 \der_3^2   \eta_1 +
        3    \der_1\eta_1 \der_2\eta_1 \der_1\eta_2 \der_1\der_3\eta_1$$ $$
+
         \der_1\eta_1 \der_2\eta_1 \der_1\eta_2 \der_2\der_3\eta_2 +
         \der_1\eta_1 \der_2\eta_1 \der_1\eta_3 \der_3^2   \eta_2 +
         \der_1\eta_1 \der_2\eta_1 \der_2\eta_2 \der_1\der_3\eta_2$$ $$ -
        3    \der_1\eta_1 \der_2\eta_1 \der_3\eta_1 \der_1^2   \eta_2 -
        3    \der_1\eta_1 \der_2\eta_1 \der_3\eta_2 \der_1\der_2\eta_2 +
         \der_1\eta_1 \der_2\eta_1 \der_3\eta_2 \der_1\der_3\eta_3 $$ $$+
         \der_1\eta_1 \der_2\eta_1 \der_3\eta_2 \der_1^2   \eta_1 -
        3    \der_1\eta_1 \der_2\eta_1 \der_3\eta_3 \der_1\der_3\eta_2 +
        2    \der_1\eta_1 \der_2\eta_2 \der_3\eta_2 \der_1\der_2\eta_1$$ $$
-
        5    \der_1\eta_1 \der_3\eta_1 \der_1\eta_2 \der_1\der_2\eta_1 +
         \der_1\eta_1 \der_3\eta_1 \der_1\eta_2 \der_2\der_3\eta_3 -
        2    \der_1\eta_1 \der_3\eta_1 \der_1\eta_3 \der_1\der_3\eta_1$$ $$
+
         \der_1\eta_1 \der_3\eta_1 \der_1\eta_3 \der_3^2   \eta_3 +
         \der_1\eta_1 \der_3\eta_1 \der_2\eta_3 \der_1\der_3\eta_2 -
        3    \der_1\eta_1 \der_3\eta_1 \der_3\eta_2 \der_1\der_2\eta_3$$ $$
-
        2    \der_1\eta_1 \der_3\eta_1 \der_3\eta_3 \der_1\der_3\eta_3 +
         \der_1\eta_1 \der_3\eta_1 \der_3\eta_3 \der_1^2   \eta_1 -
        3    \der_1\eta_1 \der_3\eta_2 \der_1\eta_3 \der_2\der_3\eta_1 $$
$$-
        2    \der_1\eta_1 \der_3\eta_2 \der_2\eta_3 \der_1\der_3\eta_1 +
        2    \der_1\eta_1 \der_3\eta_2 \der_3\eta_3 \der_1\der_2\eta_1 -
         \der_1\eta_2 \der_2\eta_2 \der_2\eta_3 \der_3^2   \eta_1$$ $$ +
        4    \der_1\eta_2 \der_2\eta_2 \der_3\eta_2 \der_2^2   \eta_1 +
        2    \der_1\eta_2 \der_2\eta_2 \der_3\eta_3 \der_2\der_3\eta_1 +
         \der_1\eta_2 \der_2\eta_3 \der_3\eta_3 \der_3^2   \eta_1 $$ $$-
        3    \der_1\eta_2 \der_3\eta_2 \der_2\eta_3 \der_2\der_3\eta_1 +
        2    \der_1\eta_2 \der_3\eta_2 \der_3\eta_3 \der_2^2   \eta_1 +
        2    \der_2\eta_1 \der_1\eta_2 \der_1\eta_3 \der_3^2   \eta_1 $$ $$-
         \der_2\eta_1 \der_1\eta_2 \der_2\eta_2 \der_1\der_3\eta_1 -
         \der_2\eta_1 \der_1\eta_2 \der_2\eta_2 \der_2\der_3\eta_2 -
        4    \der_2\eta_1 \der_1\eta_2 \der_3\eta_2 \der_1\der_2\eta_1 $$
$$-
         \der_2\eta_1 \der_1\eta_2 \der_3\eta_2 \der_2\der_3\eta_3 +
        2    \der_2\eta_1 \der_1\eta_2 \der_3\eta_2 \der_2^2   \eta_2 -
        3    \der_2\eta_1 \der_1\eta_2 \der_3\eta_3 \der_1\der_3\eta_1 $$
$$+
        2    \der_2\eta_1 \der_1\eta_2 \der_3\eta_3 \der_2\der_3\eta_2 +
        2    \der_2\eta_1 \der_1\eta_3 \der_3\eta_3 \der_3^2   \eta_2 +
         \der_2\eta_1 \der_2\eta_2 \der_1\eta_3 \der_3^2   \eta_2 $$ $$-
         \der_2\eta_1 \der_2\eta_2 \der_3\eta_2 \der_1\der_2\eta_2 -
         \der_2\eta_1 \der_2\eta_2 \der_3\eta_2 \der_1^2   \eta_1 -
         \der_2\eta_1 \der_2\eta_2 \der_3\eta_3 \der_1\der_3\eta_2 $$ $$-
        2    \der_2\eta_1 \der_3\eta_1 \der_1\eta_2 \der_1\der_2\eta_2 +
        4    \der_2\eta_1 \der_3\eta_1 \der_1\eta_2 \der_1^2   \eta_1 -
        2    \der_2\eta_1 \der_3\eta_1 \der_1\eta_3 \der_1\der_3\eta_2$$ $$
+
         \der_2\eta_1 \der_3\eta_1 \der_2\eta_2 \der_1^2   \eta_2 +
         \der_2\eta_1 \der_3\eta_1 \der_3\eta_2 \der_1^2   \eta_3 +
        2    \der_2\eta_1 \der_3\eta_2 \der_1\eta_3 \der_1\der_3\eta_1 $$
$$-
        2    \der_2\eta_1 \der_3\eta_2 \der_1\eta_3 \der_2\der_3\eta_2 +
         \der_2\eta_1 \der_3\eta_2 \der_1\eta_3 \der_3^2   \eta_3 -
         \der_2\eta_1 \der_3\eta_2 \der_3\eta_3 \der_1\der_3\eta_3$$ $$ -
         \der_2\eta_1 \der_3\eta_2 \der_3\eta_3 \der_1^2   \eta_1 +
        3    \der_2\eta_2 \der_3\eta_2 \der_1\eta_3 \der_2\der_3\eta_1 -
         \der_2\eta_2 \der_3\eta_2 \der_2\eta_3 \der_1\der_3\eta_1 $$ $$-
         \der_2\eta_2 \der_3\eta_2 \der_3\eta_3 \der_1\der_2\eta_1 -
        2    \der_3\eta_1 \der_1\eta_2 \der_1\eta_3 \der_2\der_3\eta_1 +
        3    \der_3\eta_1 \der_1\eta_2 \der_2\eta_2 \der_1\der_2\eta_1$$ $$
+
        2    \der_3\eta_1 \der_1\eta_2 \der_2\eta_3 \der_1\der_3\eta_1 -
         \der_3\eta_1 \der_1\eta_2 \der_2\eta_3 \der_2\der_3\eta_2 +
        2    \der_3\eta_1 \der_1\eta_2 \der_3\eta_2 \der_2^2   \eta_3 $$ $$-
         \der_3\eta_1 \der_1\eta_2 \der_3\eta_3 \der_1\der_2\eta_1 +
         \der_3\eta_1 \der_1\eta_2 \der_3\eta_3 \der_2\der_3\eta_3 -
         \der_3\eta_1 \der_1\eta_3 \der_2\eta_3 \der_3^2   \eta_2 $$ $$-
        2    \der_3\eta_1 \der_1\eta_3 \der_3\eta_3 \der_1\der_3\eta_1 +
         \der_3\eta_1 \der_1\eta_3 \der_3\eta_3 \der_3^2   \eta_3 -
         \der_3\eta_1 \der_2\eta_3 \der_3\eta_3 \der_1\der_3\eta_2 $$ $$-
        2    \der_3\eta_1 \der_3\eta_2 \der_1\eta_3 \der_2\der_3\eta_3 +
         \der_3\eta_1 \der_3\eta_2 \der_2\eta_3 \der_1\der_2\eta_2 +
         \der_3\eta_1 \der_3\eta_2 \der_2\eta_3 \der_1^2   \eta_1$$ $$ +
         \der_3\eta_1 \der_3\eta_2 \der_3\eta_3 \der_1\der_2\eta_3 +
        3    \der_3\eta_2 \der_1\eta_3 \der_2\eta_3 \der_3^2   \eta_1 -
        3    \der_3\eta_2 \der_1\eta_3 \der_3\eta_3 \der_2\der_3\eta_1$$ $$
+
        2    \der_3\eta_2 \der_2\eta_3 \der_3\eta_3
\der_1\der_3\eta_1)\der_1.$$

(all together $76$ terms)
\bigskip
\bigskip

$$F^{\circ 3}_{\eta_2\eta_3;\der_1}=$$
$$\eta_2\eta_3(\der_1\eta_1 \der_2\eta_1 \der_1\eta_2 \der_2\der_3\eta_1 +
         \der_1\eta_1 \der_2\eta_1 \der_1\eta_3 \der_3^2   \eta_1 +
         \der_1\eta_1 \der_2\eta_1 \der_2\eta_2 \der_1\der_3\eta_1 $$ $$+
         \der_1\eta_1 \der_2\eta_1 \der_2\eta_2 \der_2\der_3\eta_2 +
         \der_1\eta_1 \der_2\eta_1 \der_2\eta_3 \der_3^2   \eta_2 -
         \der_1\eta_1 \der_2\eta_1 \der_3\eta_1 \der_1\der_2\eta_2$$ $$ -
         \der_1\eta_1 \der_2\eta_1 \der_3\eta_1 \der_1\der_3\eta_3 -
        2    \der_1\eta_1 \der_2\eta_1 \der_3\eta_1 \der_1^2   \eta_1 -
         \der_1\eta_1 \der_2\eta_1 \der_3\eta_2 \der_1\der_2\eta_1$$ $$ -
         \der_1\eta_1 \der_2\eta_1 \der_3\eta_2 \der_2^2   \eta_2 -
         \der_1\eta_1 \der_2\eta_1 \der_3\eta_3 \der_2\der_3\eta_2 +
        2    \der_1\eta_1 \der_2\eta_2 \der_3\eta_2 \der_2^2   \eta_1$$ $$ +
        2    \der_1\eta_1 \der_2\eta_2 \der_3\eta_3 \der_2\der_3\eta_1 +
        2    \der_1\eta_1 \der_2\eta_3 \der_3\eta_3 \der_3^2   \eta_1 -
         \der_1\eta_1 \der_3\eta_1 \der_1\eta_2 \der_2^2   \eta_1$$ $$ -
         \der_1\eta_1 \der_3\eta_1 \der_1\eta_3 \der_2\der_3\eta_1 +
         \der_1\eta_1 \der_3\eta_1 \der_2\eta_2 \der_2\der_3\eta_3 +
         \der_1\eta_1 \der_3\eta_1 \der_2\eta_3 \der_1\der_3\eta_1$$ $$ +
         \der_1\eta_1 \der_3\eta_1 \der_2\eta_3 \der_3^2   \eta_3 -
         \der_1\eta_1 \der_3\eta_1 \der_3\eta_2 \der_2^2   \eta_3 -
         \der_1\eta_1 \der_3\eta_1 \der_3\eta_3 \der_1\der_2\eta_1$$ $$ -
         \der_1\eta_1 \der_3\eta_1 \der_3\eta_3 \der_2\der_3\eta_3 -
        2    \der_1\eta_1 \der_3\eta_2 \der_2\eta_3 \der_2\der_3\eta_1 -
         \der_2\eta_1 \der_1\eta_2 \der_2\eta_2 \der_2\der_3\eta_1$$ $$ +
        2    \der_2\eta_1 \der_1\eta_2 \der_3\eta_2 \der_2^2   \eta_1 +
         \der_2\eta_1 \der_1\eta_2 \der_3\eta_3 \der_2\der_3\eta_1 +
         \der_2\eta_1 \der_1\eta_3 \der_3\eta_3 \der_3^2   \eta_1$$ $$ +
         \der_2\eta_1 \der_2\eta_2 \der_1\eta_3 \der_3^2   \eta_1 +
         \der_2\eta_1 \der_2\eta_2 \der_2\eta_3 \der_3^2   \eta_2 -
        3    \der_2\eta_1 \der_2\eta_2 \der_3\eta_2 \der_1\der_2\eta_1$$ $$
-
         \der_2\eta_1 \der_2\eta_2 \der_3\eta_2 \der_2\der_3\eta_3 +
         \der_2\eta_1 \der_2\eta_2 \der_3\eta_2 \der_2^2   \eta_2 -
        2    \der_2\eta_1 \der_2\eta_2 \der_3\eta_3 \der_1\der_3\eta_1$$ $$
+
         \der_2\eta_1 \der_2\eta_2 \der_3\eta_3 \der_2\der_3\eta_2 +
        2    \der_2\eta_1 \der_2\eta_3 \der_3\eta_3 \der_3^2   \eta_2 -
        2    \der_2\eta_1 \der_3\eta_1 \der_1\eta_2 \der_1\der_2\eta_1$$ $$
-
        2    \der_2\eta_1 \der_3\eta_1 \der_1\eta_3 \der_1\der_3\eta_1 -
         \der_2\eta_1 \der_3\eta_1 \der_2\eta_2 \der_1\der_2\eta_2 +
        2    \der_2\eta_1 \der_3\eta_1 \der_2\eta_2 \der_1^2   \eta_1$$ $$ -
         \der_2\eta_1 \der_3\eta_1 \der_2\eta_3 \der_1\der_3\eta_2 -
         \der_2\eta_1 \der_3\eta_1 \der_3\eta_2 \der_1\der_2\eta_3 -
         \der_2\eta_1 \der_3\eta_1 \der_3\eta_3 \der_1\der_3\eta_3$$ $$ +
        2    \der_2\eta_1 \der_3\eta_1 \der_3\eta_3 \der_1^2   \eta_1 -
        2    \der_2\eta_1 \der_3\eta_2 \der_1\eta_3 \der_2\der_3\eta_1 +
         \der_2\eta_1 \der_3\eta_2 \der_2\eta_3 \der_1\der_3\eta_1$$ $$ -
        2    \der_2\eta_1 \der_3\eta_2 \der_2\eta_3 \der_2\der_3\eta_2 +
         \der_2\eta_1 \der_3\eta_2 \der_2\eta_3 \der_3^2   \eta_3 +
         \der_2\eta_1 \der_3\eta_2 \der_3\eta_3 \der_1\der_2\eta_1$$ $$ -
         \der_2\eta_1 \der_3\eta_2 \der_3\eta_3 \der_2\der_3\eta_3 +
         \der_2\eta_2 \der_2\eta_3 \der_3\eta_3 \der_3^2   \eta_1 -
         \der_2\eta_2 \der_3\eta_2 \der_2\eta_3 \der_2\der_3\eta_1$$ $$ +
         \der_2\eta_2 \der_3\eta_2 \der_3\eta_3 \der_2^2   \eta_1 -
         \der_3\eta_1 \der_1\eta_2 \der_2\eta_2 \der_2^2   \eta_1 -
        2    \der_3\eta_1 \der_1\eta_2 \der_2\eta_3 \der_2\der_3\eta_1$$ $$
+
         \der_3\eta_1 \der_1\eta_2 \der_3\eta_3 \der_2^2   \eta_1 -
        2    \der_3\eta_1 \der_1\eta_3 \der_2\eta_3 \der_3^2   \eta_1 +
         \der_3\eta_1 \der_1\eta_3 \der_3\eta_3 \der_2\der_3\eta_1$$ $$ +
         \der_3\eta_1 \der_2\eta_2 \der_1\eta_3 \der_2\der_3\eta_1 +
         \der_3\eta_1 \der_2\eta_2 \der_2\eta_3 \der_1\der_3\eta_1 -
         \der_3\eta_1 \der_2\eta_2 \der_2\eta_3 \der_2\der_3\eta_2$$ $$ +
        2    \der_3\eta_1 \der_2\eta_2 \der_3\eta_2 \der_2^2   \eta_3 -
        2    \der_3\eta_1 \der_2\eta_2 \der_3\eta_3 \der_1\der_2\eta_1 +
         \der_3\eta_1 \der_2\eta_2 \der_3\eta_3 \der_2\der_3\eta_3$$ $$ -
        3    \der_3\eta_1 \der_2\eta_3 \der_3\eta_3 \der_1\der_3\eta_1 -
         \der_3\eta_1 \der_2\eta_3 \der_3\eta_3 \der_2\der_3\eta_2 +
         \der_3\eta_1 \der_2\eta_3 \der_3\eta_3 \der_3^2   \eta_3$$ $$ +
         \der_3\eta_1 \der_3\eta_2 \der_2\eta_3 \der_1\der_2\eta_1 -
        2    \der_3\eta_1 \der_3\eta_2 \der_2\eta_3 \der_2\der_3\eta_3 +
         \der_3\eta_1 \der_3\eta_2 \der_2\eta_3 \der_2^2   \eta_2$$ $$ +
         \der_3\eta_1 \der_3\eta_2 \der_3\eta_3 \der_2^2   \eta_3 -
         \der_3\eta_2 \der_2\eta_3 \der_3\eta_3 \der_2\der_3\eta_1)\der_1.$$

(all together $71$ terms)

\section{\label{quadraticpart}
Quadratic differential part of $D^{10}$}

For $G\in Dif\!f_n$ denote by $G_{\eta_{i_1}\cdots\eta_{i_k};\der^{\alpha}}$
projection to subspace of
$Dif\!f_n$ generated by differential operators of the form
$\eta_{i_1}\cdots\eta_{i_k}\prod_{s,\beta\ne 0}\der^{\beta}u_s
\der^{\alpha}.$ For example, if
$$G=-5\eta_1\der_2\eta_2\der_3\eta_2\der_1\der_3^2\eta_3\der_1
+\eta_1\eta_3\der_2\eta_1\der_1\eta_3\der_1\der_2^3\eta_3\der_3^2$$ $$
-\eta_1\eta_3\der_1\eta_1\der_2\eta_1\der_2\eta_3\der_1\der_2\der_3\eta_3\der_2
+9\eta_1\eta_3\der_1\eta_1\der_2\eta_1\der_2\eta_3\der_1\der_2\der_3\eta_3\der_2$$
$$-7\eta_1\eta_3\der_1\eta_1\der_3\eta_2\der_1\eta_3\der_1^2\der_2\der_3^2\eta_3\der_3^2,$$
then
$$
G_{\eta_1\eta_3;\der_3^2}=
\eta_1\eta_3\der_2\eta_1\der_1\eta_3\der_1\der_2^3\eta_3\der_3^2
-7\eta_1\eta_3\der_1\eta_1\der_3\eta_2\der_1\eta_3\der_1^2\der_2\der_3^2\eta_3\der_3^2.$$

\begin{lm}
$F^{\circ 2}\bullet F^{\circ 3}+F\circ(F\bullet F^{\circ 3})=0.$
\end{lm}

{\bf Proof.}
Let $$Q=F^{\circ 2}\bullet F^{\circ 3}+F\circ(F\bullet F^{\circ 3}).$$
It is enough to prove that $\der_1^2$-part of $Q$ is equal to $0.$
Then by symmetry $\der^2_2$-,$\der_3^2$-parts of $Q$ should be $0,$ and
$\der_1\der_2$-, $\der_1\der_3$-, $\der_2\der_3$-parts of $G$ also will
vanish.

Let us show how to calculate $\eta_1\eta_2\eta_3 \der_1^2$-part of $Q.$

Notice that $\eta_1\eta_2\eta_3\der_1^2$-part of $F^{\circ 2}\bullet
F^{\circ 3}$, denote it by
$G_1$, is equal to
$$G_1=$$
$$F_{\eta_1;\der_1}^{\circ 2}\bullet F_{\eta_2\eta_3;\der_1}^{\circ 3}
+F_{\eta_2;\der_1}^{\circ 2}\bullet F_{\eta_1\eta_3;\der_1}^{\circ 3}
+F_{\eta_3;\der_1}^{\circ 2}\bullet F_{\eta_1\eta_2;\der_1}^{\circ 3}+
$$
$$F_{\eta_1\eta_2;\der_1}^{\circ 2}\bullet F_{\eta_3;\der_1}^{\circ 3}
+F_{\eta_1\eta_3;\der_1}^{\circ 2}\bullet F_{\eta_2;\der_1}^{\circ 3}
+F_{\eta_2\eta_3;\der_1}^{\circ 2}\bullet F_{\eta_1;\der_1}^{\circ 3}.$$

Using results of sections \ref{secondlsym}, \ref{thirdlsym} we obtain that
$$F_{\eta_1;\der_1}^{\circ 2}\bullet F_{\eta_2\eta_3;\der_1}^{\circ 3}
+F_{\eta_2;\der_1}^{\circ 2}\bullet F_{\eta_1\eta_3;\der_1}^{\circ 3}
+F_{\eta_3;\der_1}^{\circ 2}\bullet F_{\eta_1\eta_2;\der_1}^{\circ 3}
$$

$$=\eta_1\eta_2\eta_3(-5  \der_1\eta_1 \der_2\eta_1 \der_1\eta_2
\der_2\eta_2 \der_2\eta_3 \der_3\eta_3 \der_3^2 \eta_1 +
    14    \der_1\eta_1 \der_2\eta_1 \der_1\eta_2
          \der_2\eta_2 \der_3\eta_2 \der_2\eta_3 \der_2\der_3\eta_1$$ $$ -
       \der_1\eta_1 \der_2\eta_1 \der_1\eta_2 \der_2\eta_2
          \der_3\eta_2 \der_3\eta_3 \der_2^2 \eta_1 -
    8    \der_1\eta_1 \der_2\eta_1 \der_1\eta_2
          \der_3\eta_2 \der_2\eta_3 \der_3\eta_3 \der_2\der_3\eta_1$$ $$ +
    9    \der_1\eta_1 \der_2\eta_1 \der_2\eta_2
          \der_3\eta_2 \der_1\eta_3 \der_2\eta_3 \der_3^2 \eta_1 +
    4    \der_1\eta_1 \der_2\eta_1 \der_2\eta_2
          \der_3\eta_2 \der_1\eta_3 \der_3\eta_3 \der_2\der_3\eta_1$$ $$ +
    3    \der_1\eta_1 \der_2\eta_1 \der_2\eta_2
          \der_3\eta_2 \der_2\eta_3 \der_3\eta_3 \der_1\der_3\eta_1 -
    8    \der_1\eta_1 \der_2\eta_1 \der_3\eta_1
          \der_1\eta_2 \der_1\eta_3 \der_2\eta_3 \der_3^2 \eta_1$$ $$ +
    8    \der_1\eta_1 \der_2\eta_1 \der_3\eta_1
          \der_1\eta_2 \der_1\eta_3 \der_3\eta_3 \der_2\der_3\eta_1 +
    8    \der_1\eta_1 \der_2\eta_1 \der_3\eta_1
          \der_1\eta_2 \der_2\eta_2 \der_1\eta_3 \der_2\der_3\eta_1$$ $$ -
    6    \der_1\eta_1 \der_2\eta_1 \der_3\eta_1
          \der_1\eta_2 \der_2\eta_2 \der_2\eta_3 \der_2\der_3\eta_2 -
       \der_1\eta_1 \der_2\eta_1 \der_3\eta_1 \der_1\eta_2
          \der_2\eta_2 \der_2\eta_3 \der_3^2 \eta_3 $$ $$+
    10    \der_1\eta_1 \der_2\eta_1 \der_3\eta_1
          \der_1\eta_2 \der_2\eta_2 \der_3\eta_2 \der_2^2 \eta_3 -
    2    \der_1\eta_1 \der_2\eta_1 \der_3\eta_1
          \der_1\eta_2 \der_2\eta_2 \der_3\eta_3 \der_1\der_2\eta_1$$ $$ +
    4    \der_1\eta_1 \der_2\eta_1 \der_3\eta_1
          \der_1\eta_2 \der_2\eta_2 \der_3\eta_3 \der_2\der_3\eta_3 -
       \der_1\eta_1 \der_2\eta_1 \der_3\eta_1 \der_1\eta_2
          \der_2\eta_2 \der_3\eta_3 \der_2^2 \eta_2$$ $$ -
    4    \der_1\eta_1 \der_2\eta_1 \der_3\eta_1
          \der_1\eta_2 \der_2\eta_3 \der_3\eta_3 \der_1\der_3\eta_1 -
    8    \der_1\eta_1 \der_2\eta_1 \der_3\eta_1
          \der_1\eta_2 \der_2\eta_3 \der_3\eta_3 \der_2\der_3\eta_2$$ $$ +
    2    \der_1\eta_1 \der_2\eta_1 \der_3\eta_1
          \der_1\eta_2 \der_2\eta_3 \der_3\eta_3 \der_3^2 \eta_3 -
    8    \der_1\eta_1 \der_2\eta_1 \der_3\eta_1
          \der_1\eta_2 \der_3\eta_2 \der_1\eta_3 \der_2^2 \eta_1$$ $$ +
    2    \der_1\eta_1 \der_2\eta_1 \der_3\eta_1
          \der_1\eta_2 \der_3\eta_2 \der_2\eta_3 \der_1\der_2\eta_1 -
    8    \der_1\eta_1 \der_2\eta_1 \der_3\eta_1
          \der_1\eta_2 \der_3\eta_2 \der_2\eta_3 \der_2\der_3\eta_3$$ $$ +
    7    \der_1\eta_1 \der_2\eta_1 \der_3\eta_1
          \der_1\eta_2 \der_3\eta_2 \der_2\eta_3 \der_2^2 \eta_2 +
    5    \der_1\eta_1 \der_2\eta_1 \der_3\eta_1
          \der_1\eta_2 \der_3\eta_2 \der_3\eta_3 \der_2^2 \eta_3$$ $$ -
    10    \der_1\eta_1 \der_2\eta_1 \der_3\eta_1
          \der_1\eta_3 \der_2\eta_3 \der_3\eta_3 \der_3^2 \eta_2 +
    5    \der_1\eta_1 \der_2\eta_1 \der_3\eta_1
          \der_2\eta_2 \der_1\eta_3 \der_2\eta_3 \der_3^2 \eta_2$$ $$ -
    2    \der_1\eta_1 \der_2\eta_1 \der_3\eta_1
          \der_2\eta_2 \der_1\eta_3 \der_3\eta_3 \der_1\der_3\eta_1 +
    4    \der_1\eta_1 \der_2\eta_1 \der_3\eta_1
          \der_2\eta_2 \der_1\eta_3 \der_3\eta_3 \der_2\der_3\eta_2$$ $$ -
       \der_1\eta_1 \der_2\eta_1 \der_3\eta_1 \der_2\eta_2
          \der_1\eta_3 \der_3\eta_3 \der_3^2 \eta_3 +
    5    \der_1\eta_1 \der_2\eta_1 \der_3\eta_1
          \der_2\eta_2 \der_2\eta_3 \der_3\eta_3 \der_1\der_3\eta_2$$ $$ +
    4    \der_1\eta_1 \der_2\eta_1 \der_3\eta_1
          \der_2\eta_2 \der_3\eta_2 \der_1\eta_3 \der_1\der_2\eta_1 +
    8    \der_1\eta_1 \der_2\eta_1 \der_3\eta_1
          \der_2\eta_2 \der_3\eta_2 \der_1\eta_3 \der_2\der_3\eta_3$$ $$ -
    2    \der_1\eta_1 \der_2\eta_1 \der_3\eta_1
          \der_2\eta_2 \der_3\eta_2 \der_1\eta_3 \der_2^2 \eta_2 -
    7  \der_1\eta_1 \der_2\eta_1 \der_3\eta_1 \der_2\eta_2
          \der_3\eta_2 \der_2\eta_3 \der_1\der_2\eta_2$$ $$ -
    2    \der_1\eta_1 \der_2\eta_1 \der_3\eta_1
          \der_2\eta_2 \der_3\eta_2 \der_2\eta_3 \der_1\der_3\eta_3 -
    5    \der_1\eta_1 \der_2\eta_1 \der_3\eta_1
          \der_2\eta_2 \der_3\eta_2 \der_3\eta_3 \der_1\der_2\eta_3$$ $$ +
    2    \der_1\eta_1 \der_2\eta_1 \der_3\eta_1
          \der_3\eta_2 \der_1\eta_3 \der_2\eta_3 \der_1\der_3\eta_1 -
    8    \der_1\eta_1 \der_2\eta_1 \der_3\eta_1
          \der_3\eta_2 \der_1\eta_3 \der_2\eta_3 \der_2\der_3\eta_2$$ $$ +
    7    \der_1\eta_1 \der_2\eta_1 \der_3\eta_1
          \der_3\eta_2 \der_1\eta_3 \der_2\eta_3 \der_3^2 \eta_3 -
    6    \der_1\eta_1 \der_2\eta_1 \der_3\eta_1
          \der_3\eta_2 \der_1\eta_3 \der_3\eta_3 \der_2\der_3\eta_3$$ $$ -
       \der_1\eta_1 \der_2\eta_1 \der_3\eta_1 \der_3\eta_2
          \der_1\eta_3 \der_3\eta_3 \der_2^2 \eta_2 +
    2    \der_1\eta_1 \der_2\eta_1 \der_3\eta_1
          \der_3\eta_2 \der_2\eta_3 \der_3\eta_3 \der_1\der_2\eta_2$$ $$ +
    7    \der_1\eta_1 \der_2\eta_1 \der_3\eta_1
          \der_3\eta_2 \der_2\eta_3 \der_3\eta_3 \der_1\der_3\eta_3 +
    13    \der_1\eta_1 \der_2\eta_1 \der_3\eta_2
          \der_1\eta_3 \der_2\eta_3 \der_3\eta_3 \der_3^2 \eta_1$$ $$ +
    4    \der_1\eta_1 \der_3\eta_1 \der_1\eta_2
          \der_2\eta_2 \der_2\eta_3 \der_3\eta_3 \der_2\der_3\eta_1 -
    13    \der_1\eta_1 \der_3\eta_1 \der_1\eta_2
          \der_2\eta_2 \der_3\eta_2 \der_2\eta_3 \der_2^2 \eta_1$$ $$ +
    9    \der_1\eta_1 \der_3\eta_1 \der_1\eta_2
          \der_3\eta_2 \der_2\eta_3 \der_3\eta_3 \der_2^2 \eta_1 +
       \der_1\eta_1 \der_3\eta_1 \der_2\eta_2 \der_1\eta_3
          \der_2\eta_3 \der_3\eta_3 \der_3^2 \eta_1$$ $$ -
    8    \der_1\eta_1 \der_3\eta_1 \der_2\eta_2
          \der_3\eta_2 \der_1\eta_3 \der_2\eta_3 \der_2\der_3\eta_1 -
    5    \der_1\eta_1 \der_3\eta_1 \der_2\eta_2
          \der_3\eta_2 \der_1\eta_3 \der_3\eta_3 \der_2^2 \eta_1$$ $$ -
    3    \der_1\eta_1 \der_3\eta_1 \der_2\eta_2
          \der_3\eta_2 \der_2\eta_3 \der_3\eta_3 \der_1\der_2\eta_1 -
    14    \der_1\eta_1 \der_3\eta_1 \der_3\eta_2
          \der_1\eta_3 \der_2\eta_3 \der_3\eta_3 \der_2\der_3\eta_1$$ $$ +
       \der_2\eta_1 \der_1\eta_2 \der_2\eta_2 \der_3\eta_2
          \der_2\eta_3 \der_3\eta_3 \der_2\der_3\eta_1 +
       \der_2\eta_1 \der_2\eta_2 \der_3\eta_2 \der_1\eta_3
          \der_2\eta_3 \der_3\eta_3 \der_3^2 \eta_1 $$ $$-
    2    \der_2\eta_1 \der_3\eta_1 \der_1\eta_2
          \der_1\eta_3 \der_2\eta_3 \der_3\eta_3 \der_3^2 \eta_1 -
    4    \der_2\eta_1 \der_3\eta_1 \der_1\eta_2
          \der_2\eta_2 \der_1\eta_3 \der_2\eta_3 \der_3^2 \eta_1$$ $$ +
    6    \der_2\eta_1 \der_3\eta_1 \der_1\eta_2
          \der_2\eta_2 \der_1\eta_3 \der_3\eta_3 \der_2\der_3\eta_1 -
    8    \der_2\eta_1 \der_3\eta_1 \der_1\eta_2
          \der_2\eta_2 \der_2\eta_3 \der_3\eta_3 \der_1\der_3\eta_1$$ $$ +
       \der_2\eta_1 \der_3\eta_1 \der_1\eta_2 \der_2\eta_2
          \der_2\eta_3 \der_3\eta_3 \der_2\der_3\eta_2 +
       \der_2\eta_1 \der_3\eta_1 \der_1\eta_2 \der_2\eta_2
          \der_2\eta_3 \der_3\eta_3 \der_3^2 \eta_3$$ $$ -
    2    \der_2\eta_1 \der_3\eta_1 \der_1\eta_2
          \der_2\eta_2 \der_3\eta_2 \der_1\eta_3 \der_2^2 \eta_1 +
    13    \der_2\eta_1 \der_3\eta_1 \der_1\eta_2
          \der_2\eta_2 \der_3\eta_2 \der_2\eta_3 \der_1\der_2\eta_1$$ $$ -
    2    \der_2\eta_1 \der_3\eta_1 \der_1\eta_2
          \der_2\eta_2 \der_3\eta_2 \der_2\eta_3 \der_2\der_3\eta_3 -
    2    \der_2\eta_1 \der_3\eta_1 \der_1\eta_2
          \der_2\eta_2 \der_3\eta_2 \der_2\eta_3 \der_2^2 \eta_2$$ $$ +
    2    \der_2\eta_1 \der_3\eta_1 \der_1\eta_2
          \der_3\eta_2 \der_1\eta_3 \der_2\eta_3 \der_2\der_3\eta_1 -
    4    \der_2\eta_1 \der_3\eta_1 \der_1\eta_2
          \der_3\eta_2 \der_1\eta_3 \der_3\eta_3 \der_2^2 \eta_1$$ $$ -
    5    \der_2\eta_1 \der_3\eta_1 \der_1\eta_2
          \der_3\eta_2 \der_2\eta_3 \der_3\eta_3 \der_1\der_2\eta_1 +
       \der_2\eta_1 \der_3\eta_1 \der_1\eta_2 \der_3\eta_2
          \der_2\eta_3 \der_3\eta_3 \der_2\der_3\eta_3$$ $$ +
       \der_2\eta_1 \der_3\eta_1 \der_1\eta_2 \der_3\eta_2
          \der_2\eta_3 \der_3\eta_3 \der_2^2 \eta_2+
    5    \der_2\eta_1 \der_3\eta_1 \der_2\eta_2
          \der_3\eta_2 \der_1\eta_3 \der_2\eta_3 \der_1\der_3\eta_1$$ $$ -
       \der_2\eta_1 \der_3\eta_1 \der_2\eta_2 \der_3\eta_2
          \der_1\eta_3 \der_2\eta_3 \der_2\der_3\eta_2 -
       \der_2\eta_1 \der_3\eta_1 \der_2\eta_2 \der_3\eta_2
          \der_1\eta_3 \der_2\eta_3 \der_3^2 \eta_3$$ $$ +
    8    \der_2\eta_1 \der_3\eta_1 \der_2\eta_2
          \der_3\eta_2 \der_1\eta_3 \der_3\eta_3 \der_1\der_2\eta_1 -
       \der_2\eta_1 \der_3\eta_1 \der_2\eta_2 \der_3\eta_2
          \der_1\eta_3 \der_3\eta_3 \der_2\der_3\eta_3$$ $$ -
       \der_2\eta_1 \der_3\eta_1 \der_2\eta_2 \der_3\eta_2
          \der_1\eta_3 \der_3\eta_3 \der_2^2 \eta_2+
    13    \der_2\eta_1 \der_3\eta_1 \der_3\eta_2
          \der_1\eta_3 \der_2\eta_3 \der_3\eta_3 \der_1\der_3\eta_1$$ $$ -
    2    \der_2\eta_1 \der_3\eta_1 \der_3\eta_2
          \der_1\eta_3 \der_2\eta_3 \der_3\eta_3 \der_2\der_3\eta_2 -
    2    \der_2\eta_1 \der_3\eta_1 \der_3\eta_2
          \der_1\eta_3 \der_2\eta_3 \der_3\eta_3 \der_3^2 \eta_3$$ $$ -
       \der_3\eta_1 \der_1\eta_2 \der_2\eta_2 \der_3\eta_2
          \der_2\eta_3 \der_3\eta_3 \der_2^2 \eta_1 -
       \der_3\eta_1 \der_2\eta_2 \der_3\eta_2 \der_1\eta_3
          \der_2\eta_3 \der_3\eta_3 \der_2\der_3\eta_1)\der_1^2.$$
Similarly,
$$F_{\eta_1\eta_2;\der_1}^{\circ 2}\bullet F_{\eta_3;\der_1}^{\circ 3}
+F_{\eta_1\eta_3;\der_1}^{\circ 2}\bullet F_{\eta_2;\der_1}^{\circ 3}
+F_{\eta_2\eta_3;\der_1}^{\circ 2}\bullet F_{\eta_1;\der_1}^{\circ 3}=$$

$$\eta_1\eta_2\eta_3(-4 \der_1\eta_1 \der_2\eta_1 \der_1\eta_2 \der_2\eta_2
          \der_3\eta_2 \der_2\eta_3 \der_2\der_3\eta_1 +
       \der_1\eta_1 \der_2\eta_1 \der_1\eta_2 \der_2\eta_2
          \der_3\eta_2 \der_3\eta_3 \der_2^2 \eta_1 $$ $$+
    3    \der_1\eta_1 \der_2\eta_1 \der_1\eta_2
          \der_3\eta_2 \der_2\eta_3 \der_3\eta_3 \der_2\der_3\eta_1 -
    4    \der_1\eta_1 \der_2\eta_1 \der_2\eta_2
          \der_3\eta_2 \der_1\eta_3 \der_2\eta_3 \der_3^2 \eta_1$$ $$ +
       \der_1\eta_1 \der_2\eta_1 \der_2\eta_2 \der_3\eta_2
          \der_1\eta_3 \der_3\eta_3 \der_2\der_3\eta_1 -
    3    \der_1\eta_1 \der_2\eta_1 \der_2\eta_2
          \der_3\eta_2 \der_2\eta_3 \der_3\eta_3 \der_1\der_3\eta_1$$ $$ +
    8    \der_1\eta_1 \der_2\eta_1 \der_3\eta_1
          \der_1\eta_2 \der_1\eta_3 \der_2\eta_3 \der_3^2 \eta_1 -
    8    \der_1\eta_1 \der_2\eta_1 \der_3\eta_1
          \der_1\eta_2 \der_1\eta_3 \der_3\eta_3 \der_2\der_3\eta_1$$ $$ -
    8    \der_1\eta_1 \der_2\eta_1 \der_3\eta_1
          \der_1\eta_2 \der_2\eta_2 \der_1\eta_3 \der_2\der_3\eta_1 +
    4    \der_1\eta_1 \der_2\eta_1 \der_3\eta_1
          \der_1\eta_2 \der_2\eta_2 \der_2\eta_3 \der_1\der_3\eta_1$$ $$ +
       \der_1\eta_1 \der_2\eta_1 \der_3\eta_1 \der_1\eta_2
          \der_2\eta_2 \der_3\eta_3 \der_1\der_2\eta_1 +
    6    \der_1\eta_1 \der_2\eta_1 \der_3\eta_1
          \der_1\eta_2 \der_2\eta_3 \der_3\eta_3 \der_1\der_3\eta_1$$ $$ +
    8    \der_1\eta_1 \der_2\eta_1 \der_3\eta_1
          \der_1\eta_2 \der_3\eta_2 \der_1\eta_3 \der_2^2 \eta_1 -
    5    \der_1\eta_1 \der_2\eta_1 \der_3\eta_1
          \der_1\eta_2 \der_3\eta_2 \der_2\eta_3 \der_1\der_2\eta_1$$ $$ +
       \der_1\eta_1 \der_2\eta_1 \der_3\eta_1 \der_2\eta_2
          \der_1\eta_3 \der_3\eta_3 \der_1\der_3\eta_1-
    6    \der_1\eta_1 \der_2\eta_1 \der_3\eta_1
          \der_2\eta_2 \der_3\eta_2 \der_1\eta_3 \der_1\der_2\eta_1$$ $$ +
    3    \der_1\eta_1 \der_2\eta_1 \der_3\eta_1
          \der_2\eta_2 \der_3\eta_2 \der_2\eta_3 \der_1^2 \eta_1 -
    5    \der_1\eta_1 \der_2\eta_1 \der_3\eta_1
          \der_3\eta_2 \der_1\eta_3 \der_2\eta_3 \der_1\der_3\eta_1$$ $$ +
    4    \der_1\eta_1 \der_2\eta_1 \der_3\eta_1
          \der_3\eta_2 \der_1\eta_3 \der_3\eta_3 \der_1\der_2\eta_1 -
    3    \der_1\eta_1 \der_2\eta_1 \der_3\eta_1
          \der_3\eta_2 \der_2\eta_3 \der_3\eta_3 \der_1^2 \eta_1$$ $$ -
    3    \der_1\eta_1 \der_2\eta_1 \der_3\eta_2
          \der_1\eta_3 \der_2\eta_3 \der_3\eta_3 \der_3^2 \eta_1 +
       \der_1\eta_1 \der_3\eta_1 \der_1\eta_2 \der_2\eta_2
          \der_2\eta_3 \der_3\eta_3 \der_2\der_3\eta_1$$ $$ +
    3    \der_1\eta_1 \der_3\eta_1 \der_1\eta_2
          \der_2\eta_2 \der_3\eta_2 \der_2\eta_3 \der_2^2 \eta_1 -
    4    \der_1\eta_1 \der_3\eta_1 \der_1\eta_2
          \der_3\eta_2 \der_2\eta_3 \der_3\eta_3 \der_2^2 \eta_1$$ $$ -
       \der_1\eta_1 \der_3\eta_1 \der_2\eta_2 \der_1\eta_3
          \der_2\eta_3 \der_3\eta_3 \der_3^2 \eta_1 +
    3    \der_1\eta_1 \der_3\eta_1 \der_2\eta_2
          \der_3\eta_2 \der_1\eta_3 \der_2\eta_3 \der_2\der_3\eta_1$$ $$ +
    3    \der_1\eta_1 \der_3\eta_1 \der_2\eta_2
          \der_3\eta_2 \der_2\eta_3 \der_3\eta_3 \der_1\der_2\eta_1 +
    4    \der_1\eta_1 \der_3\eta_1 \der_3\eta_2
          \der_1\eta_3 \der_2\eta_3 \der_3\eta_3 \der_2\der_3\eta_1$$ $$ -
       \der_2\eta_1 \der_1\eta_2 \der_2\eta_2 \der_3\eta_2
          \der_2\eta_3 \der_3\eta_3 \der_2\der_3\eta_1 -
       \der_2\eta_1 \der_2\eta_2 \der_3\eta_2 \der_1\eta_3
          \der_2\eta_3 \der_3\eta_3 \der_3^2 \eta_1$$ $$ +
    2    \der_2\eta_1 \der_3\eta_1 \der_1\eta_2
          \der_1\eta_3 \der_2\eta_3 \der_3\eta_3 \der_3^2 \eta_1 +
    4    \der_2\eta_1 \der_3\eta_1 \der_1\eta_2
          \der_2\eta_2 \der_1\eta_3 \der_2\eta_3 \der_3^2 \eta_1$$ $$ -
    6    \der_2\eta_1 \der_3\eta_1 \der_1\eta_2
          \der_2\eta_2 \der_1\eta_3 \der_3\eta_3 \der_2\der_3\eta_1 +
    4    \der_2\eta_1 \der_3\eta_1 \der_1\eta_2
          \der_2\eta_2 \der_2\eta_3 \der_3\eta_3 \der_1\der_3\eta_1$$ $$ +
    2    \der_2\eta_1 \der_3\eta_1 \der_1\eta_2
          \der_2\eta_2 \der_3\eta_2 \der_1\eta_3 \der_2^2 \eta_1 -
    5    \der_2\eta_1 \der_3\eta_1 \der_1\eta_2
          \der_2\eta_2 \der_3\eta_2 \der_2\eta_3 \der_1\der_2\eta_1$$ $$ -
    2    \der_2\eta_1 \der_3\eta_1 \der_1\eta_2
          \der_3\eta_2 \der_1\eta_3 \der_2\eta_3 \der_2\der_3\eta_1 +
    4    \der_2\eta_1 \der_3\eta_1 \der_1\eta_2
          \der_3\eta_2 \der_1\eta_3 \der_3\eta_3 \der_2^2 \eta_1$$ $$ +
       \der_2\eta_1 \der_3\eta_1 \der_1\eta_2 \der_3\eta_2
          \der_2\eta_3 \der_3\eta_3 \der_1\der_2\eta_1 -
       \der_2\eta_1 \der_3\eta_1 \der_2\eta_2 \der_3\eta_2
          \der_1\eta_3 \der_2\eta_3 \der_1\der_3\eta_1$$ $$ -
    4    \der_2\eta_1 \der_3\eta_1 \der_2\eta_2
          \der_3\eta_2 \der_1\eta_3 \der_3\eta_3 \der_1\der_2\eta_1 -
    5    \der_2\eta_1 \der_3\eta_1 \der_3\eta_2
          \der_1\eta_3 \der_2\eta_3 \der_3\eta_3 \der_1\der_3\eta_1$$ $$ +
       \der_3\eta_1 \der_1\eta_2 \der_2\eta_2 \der_3\eta_2
          \der_2\eta_3 \der_3\eta_3 \der_2^2 \eta_1 +
       \der_3\eta_1 \der_2\eta_2 \der_3\eta_2 \der_1\eta_3
          \der_2\eta_3 \der_3\eta_3 \der_2\der_3\eta_1)\der_1^2.$$
Thus,
$$G_1=$$
$$\eta_1\eta_2\eta_3(-5  \der_1\eta_1 \der_2\eta_1 \der_1\eta_2 \der_2\eta_2
          \der_2\eta_3 \der_3\eta_3 \der_3^2 \eta_1 +
    10    \der_1\eta_1 \der_2\eta_1 \der_1\eta_2
          \der_2\eta_2 \der_3\eta_2 \der_2\eta_3 \der_2\der_3\eta_1$$ $$ -
    5    \der_1\eta_1 \der_2\eta_1 \der_1\eta_2
          \der_3\eta_2 \der_2\eta_3 \der_3\eta_3 \der_2\der_3\eta_1 +
    5    \der_1\eta_1 \der_2\eta_1 \der_2\eta_2
          \der_3\eta_2 \der_1\eta_3 \der_2\eta_3 \der_3^2 \eta_1$$ $$ +
    5    \der_1\eta_1 \der_2\eta_1 \der_2\eta_2
          \der_3\eta_2 \der_1\eta_3 \der_3\eta_3 \der_2\der_3\eta_1 +
    4    \der_1\eta_1 \der_2\eta_1 \der_3\eta_1
          \der_1\eta_2 \der_2\eta_2 \der_2\eta_3 \der_1\der_3\eta_1$$ $$ -
    6    \der_1\eta_1 \der_2\eta_1 \der_3\eta_1
          \der_1\eta_2 \der_2\eta_2 \der_2\eta_3 \der_2\der_3\eta_2 -
       \der_1\eta_1 \der_2\eta_1 \der_3\eta_1 \der_1\eta_2
          \der_2\eta_2 \der_2\eta_3 \der_3^2 \eta_3$$ $$ +
    10    \der_1\eta_1 \der_2\eta_1 \der_3\eta_1
          \der_1\eta_2 \der_2\eta_2 \der_3\eta_2 \der_2^2 \eta_3 -
       \der_1\eta_1 \der_2\eta_1 \der_3\eta_1 \der_1\eta_2
          \der_2\eta_2 \der_3\eta_3 \der_1\der_2\eta_1$$ $$ +
    4    \der_1\eta_1 \der_2\eta_1 \der_3\eta_1
          \der_1\eta_2 \der_2\eta_2 \der_3\eta_3 \der_2\der_3\eta_3 -
       \der_1\eta_1 \der_2\eta_1 \der_3\eta_1 \der_1\eta_2
          \der_2\eta_2 \der_3\eta_3 \der_2^2 \eta_2$$ $$ +
    2    \der_1\eta_1 \der_2\eta_1 \der_3\eta_1
          \der_1\eta_2 \der_2\eta_3 \der_3\eta_3 \der_1\der_3\eta_1 -
    8    \der_1\eta_1 \der_2\eta_1 \der_3\eta_1
          \der_1\eta_2 \der_2\eta_3 \der_3\eta_3 \der_2\der_3\eta_2$$ $$ +
    2    \der_1\eta_1 \der_2\eta_1 \der_3\eta_1
          \der_1\eta_2 \der_2\eta_3 \der_3\eta_3 \der_3^2 \eta_3-
    3    \der_1\eta_1 \der_2\eta_1 \der_3\eta_1
          \der_1\eta_2 \der_3\eta_2 \der_2\eta_3 \der_1\der_2\eta_1$$ $$ -
    8    \der_1\eta_1 \der_2\eta_1 \der_3\eta_1
          \der_1\eta_2 \der_3\eta_2 \der_2\eta_3 \der_2\der_3\eta_3 +
    7    \der_1\eta_1 \der_2\eta_1 \der_3\eta_1
          \der_1\eta_2 \der_3\eta_2 \der_2\eta_3 \der_2^2 \eta_2$$ $$ +
    5    \der_1\eta_1 \der_2\eta_1 \der_3\eta_1
          \der_1\eta_2 \der_3\eta_2 \der_3\eta_3 \der_2^2 \eta_3 -
    10    \der_1\eta_1 \der_2\eta_1 \der_3\eta_1
          \der_1\eta_3 \der_2\eta_3 \der_3\eta_3 \der_3^2 \eta_2 $$ $$+
    5    \der_1\eta_1 \der_2\eta_1 \der_3\eta_1
          \der_2\eta_2 \der_1\eta_3 \der_2\eta_3 \der_3^2 \eta_2 -
       \der_1\eta_1 \der_2\eta_1 \der_3\eta_1 \der_2\eta_2
          \der_1\eta_3 \der_3\eta_3 \der_1\der_3\eta_1$$ $$ +
    4    \der_1\eta_1 \der_2\eta_1 \der_3\eta_1
          \der_2\eta_2 \der_1\eta_3 \der_3\eta_3 \der_2\der_3\eta_2 -
       \der_1\eta_1 \der_2\eta_1 \der_3\eta_1 \der_2\eta_2
          \der_1\eta_3 \der_3\eta_3 \der_3^2 \eta_3$$ $$ +
    5    \der_1\eta_1 \der_2\eta_1 \der_3\eta_1
          \der_2\eta_2 \der_2\eta_3 \der_3\eta_3 \der_1\der_3\eta_2 -
    2    \der_1\eta_1 \der_2\eta_1 \der_3\eta_1
          \der_2\eta_2 \der_3\eta_2 \der_1\eta_3 \der_1\der_2\eta_1$$ $$ +
    8    \der_1\eta_1 \der_2\eta_1 \der_3\eta_1
          \der_2\eta_2 \der_3\eta_2 \der_1\eta_3 \der_2\der_3\eta_3 -
    2    \der_1\eta_1 \der_2\eta_1 \der_3\eta_1
          \der_2\eta_2 \der_3\eta_2 \der_1\eta_3 \der_2^2 \eta_2$$ $$ -
    7    \der_1\eta_1 \der_2\eta_1 \der_3\eta_1
          \der_2\eta_2 \der_3\eta_2 \der_2\eta_3 \der_1\der_2\eta_2 -
    2    \der_1\eta_1 \der_2\eta_1 \der_3\eta_1
          \der_2\eta_2 \der_3\eta_2 \der_2\eta_3 \der_1\der_3\eta_3$$ $$ +
    3    \der_1\eta_1 \der_2\eta_1 \der_3\eta_1
          \der_2\eta_2 \der_3\eta_2 \der_2\eta_3 \der_1^2 \eta_1 -
    5    \der_1\eta_1 \der_2\eta_1 \der_3\eta_1
          \der_2\eta_2 \der_3\eta_2 \der_3\eta_3 \der_1\der_2\eta_3$$ $$ -
    3    \der_1\eta_1 \der_2\eta_1 \der_3\eta_1
          \der_3\eta_2 \der_1\eta_3 \der_2\eta_3 \der_1\der_3\eta_1 -
    8    \der_1\eta_1 \der_2\eta_1 \der_3\eta_1
          \der_3\eta_2 \der_1\eta_3 \der_2\eta_3 \der_2\der_3\eta_2$$ $$ +
    7    \der_1\eta_1 \der_2\eta_1 \der_3\eta_1
          \der_3\eta_2 \der_1\eta_3 \der_2\eta_3 \der_3^2 \eta_3 +
    4    \der_1\eta_1 \der_2\eta_1 \der_3\eta_1
          \der_3\eta_2 \der_1\eta_3 \der_3\eta_3 \der_1\der_2\eta_1$$ $$ -
    6    \der_1\eta_1 \der_2\eta_1 \der_3\eta_1
          \der_3\eta_2 \der_1\eta_3 \der_3\eta_3 \der_2\der_3\eta_3 -
       \der_1\eta_1 \der_2\eta_1 \der_3\eta_1 \der_3\eta_2
          \der_1\eta_3 \der_3\eta_3 \der_2^2 \eta_2$$ $$ +
    2    \der_1\eta_1 \der_2\eta_1 \der_3\eta_1
          \der_3\eta_2 \der_2\eta_3 \der_3\eta_3 \der_1\der_2\eta_2+
    7    \der_1\eta_1 \der_2\eta_1 \der_3\eta_1
          \der_3\eta_2 \der_2\eta_3 \der_3\eta_3 \der_1\der_3\eta_3$$ $$ -
    3    \der_1\eta_1 \der_2\eta_1 \der_3\eta_1
          \der_3\eta_2 \der_2\eta_3 \der_3\eta_3 \der_1^2 \eta_1 +
    10    \der_1\eta_1 \der_2\eta_1 \der_3\eta_2
          \der_1\eta_3 \der_2\eta_3 \der_3\eta_3 \der_3^2 \eta_1$$ $$ +
    5    \der_1\eta_1 \der_3\eta_1 \der_1\eta_2
          \der_2\eta_2 \der_2\eta_3 \der_3\eta_3 \der_2\der_3\eta_1 -
    10    \der_1\eta_1 \der_3\eta_1 \der_1\eta_2
          \der_2\eta_2 \der_3\eta_2 \der_2\eta_3 \der_2^2 \eta_1$$ $$ +
    5    \der_1\eta_1 \der_3\eta_1 \der_1\eta_2
          \der_3\eta_2 \der_2\eta_3 \der_3\eta_3 \der_2^2 \eta_1 -
    5    \der_1\eta_1 \der_3\eta_1 \der_2\eta_2
          \der_3\eta_2 \der_1\eta_3 \der_2\eta_3 \der_2\der_3\eta_1$$ $$ -
    5    \der_1\eta_1 \der_3\eta_1 \der_2\eta_2
          \der_3\eta_2 \der_1\eta_3 \der_3\eta_3 \der_2^2 \eta_1 -
    10    \der_1\eta_1 \der_3\eta_1 \der_3\eta_2
          \der_1\eta_3 \der_2\eta_3 \der_3\eta_3 \der_2\der_3\eta_1$$ $$ -
    4    \der_2\eta_1 \der_3\eta_1 \der_1\eta_2
          \der_2\eta_2 \der_2\eta_3 \der_3\eta_3 \der_1\der_3\eta_1 +
       \der_2\eta_1 \der_3\eta_1 \der_1\eta_2 \der_2\eta_2
          \der_2\eta_3 \der_3\eta_3 \der_2\der_3\eta_2$$ $$ +
       \der_2\eta_1 \der_3\eta_1 \der_1\eta_2 \der_2\eta_2
          \der_2\eta_3 \der_3\eta_3 \der_3^2 \eta_3+
    8    \der_2\eta_1 \der_3\eta_1 \der_1\eta_2
          \der_2\eta_2 \der_3\eta_2 \der_2\eta_3 \der_1\der_2\eta_1$$ $$ -
    2    \der_2\eta_1 \der_3\eta_1 \der_1\eta_2
          \der_2\eta_2 \der_3\eta_2 \der_2\eta_3 \der_2\der_3\eta_3 -
    2    \der_2\eta_1 \der_3\eta_1 \der_1\eta_2
          \der_2\eta_2 \der_3\eta_2 \der_2\eta_3 \der_2^2 \eta_2 $$ $$-
    4    \der_2\eta_1 \der_3\eta_1 \der_1\eta_2
          \der_3\eta_2 \der_2\eta_3 \der_3\eta_3 \der_1\der_2\eta_1 +
       \der_2\eta_1 \der_3\eta_1 \der_1\eta_2 \der_3\eta_2
          \der_2\eta_3 \der_3\eta_3 \der_2\der_3\eta_3$$ $$ +
       \der_2\eta_1 \der_3\eta_1 \der_1\eta_2 \der_3\eta_2
          \der_2\eta_3 \der_3\eta_3 \der_2^2 \eta_2+
    4    \der_2\eta_1 \der_3\eta_1 \der_2\eta_2
          \der_3\eta_2 \der_1\eta_3 \der_2\eta_3 \der_1\der_3\eta_1$$ $$ -
       \der_2\eta_1 \der_3\eta_1 \der_2\eta_2 \der_3\eta_2
          \der_1\eta_3 \der_2\eta_3 \der_2\der_3\eta_2 -
       \der_2\eta_1 \der_3\eta_1 \der_2\eta_2 \der_3\eta_2
          \der_1\eta_3 \der_2\eta_3 \der_3^2 \eta_3 $$ $$+
    4    \der_2\eta_1 \der_3\eta_1 \der_2\eta_2
          \der_3\eta_2 \der_1\eta_3 \der_3\eta_3 \der_1\der_2\eta_1 -
       \der_2\eta_1 \der_3\eta_1 \der_2\eta_2 \der_3\eta_2
          \der_1\eta_3 \der_3\eta_3 \der_2\der_3\eta_3$$ $$ -
       \der_2\eta_1 \der_3\eta_1 \der_2\eta_2 \der_3\eta_2
          \der_1\eta_3 \der_3\eta_3 \der_2^2 \eta_2+
    8    \der_2\eta_1 \der_3\eta_1 \der_3\eta_2
          \der_1\eta_3 \der_2\eta_3 \der_3\eta_3 \der_1\der_3\eta_1$$ $$ -
    2    \der_2\eta_1 \der_3\eta_1 \der_3\eta_2
          \der_1\eta_3 \der_2\eta_3 \der_3\eta_3 \der_2\der_3\eta_2 -
    2    \der_2\eta_1 \der_3\eta_1 \der_3\eta_2
          \der_1\eta_3 \der_2\eta_3 \der_3\eta_3 \der_3^2 \eta_3)\der_1^2.$$

Now calculate $\eta_1\eta_2\eta_3\der_1^2$-part of $F\circ(F\bullet F^{\circ
3}).$ Set $G=F\bullet F^{\circ 3}.$
It is easy to see that
$$G_{\eta_1\eta_2;\der_1^2}=F_{\eta_1;\der_1}\bullet F^{\circ
3}_{\eta_2;\der_1}+F_{\eta_2;\der_1}\bullet F^{\circ 3}_{\eta_1;\der_1},$$

$$G_{\eta_1\eta_3;\der_1^2}=F_{\eta_1;\der_1}\bullet F^{\circ
3}_{\eta_3;\der_1}+F_{\eta_3;\der_1}\bullet F^{\circ 3}_{\eta_1;\der_1},$$

$$G_{\eta_2\eta_3;\der_1^2}=F_{\eta_2;\der_1}\bullet F^{\circ
3}_{\eta_3;\der_1}+F_{\eta_3;\der_1}\bullet F^{\circ 3}_{\eta_2;\der_1}.$$
By results of section \ref{thirdlsym},
$$G_{\eta_1\eta_2;\der_1^2}=\eta_1\eta_2 H_{\eta_1\eta_2} \der_1^2,$$
$$G_{\eta_1\eta_3;\der_1^2}=\eta_1\eta_3 H_{\eta_1\eta_3} \der_1^2,$$
$$G_{\eta_2\eta_3;\der_1^2}=\eta_2\eta_3 H_{\eta_2\eta_3} \der_1^2,$$
where,
$$H_{\eta_1\eta_2}=$$
$$4\der_1\eta_1 \der_1\eta_2 \der_2\eta_1 \der_2\eta_2 \der_2\eta_3
\der_3\eta_1 -
  2\der_1\eta_1 \der_1\eta_2 \der_2\eta_1 \der_2\eta_3 \der_3\eta_1
        \der_3\eta_3 $$ $$-
   \der_1\eta_1 \der_1\eta_3 \der_2\eta_1 \der_2\eta_2 \der_3\eta_1
      \der_3\eta_3 +
  3 \der_1\eta_1 \der_1\eta_3 \der_2\eta_1 \der_2\eta_3 \der_3\eta_1
        \der_3\eta_2 $$ $$+
  \der_1\eta_2 \der_2\eta_1 \der_2\eta_2 \der_2\eta_3 \der_3\eta_1
      \der_3\eta_3 +
  \der_1\eta_3 \der_2\eta_1 \der_2\eta_2 \der_2\eta_3 \der_3\eta_1
      \der_3\eta_2$$ $$ +
  2 \der_1\eta_3 \der_2\eta_1 \der_2\eta_3 \der_3\eta_1 \der_3\eta_2
        \der_3\eta_3,$$

$$H_{\eta_1\eta_3}=$$
$$- \der_1\eta_1 \der_1\eta_2 \der_2\eta_1 \der_2\eta_2 \der_3\eta_1
\der_3\eta_3 +
  3 \der_1\eta_1 \der_1\eta_2 \der_2\eta_1 \der_2\eta_3 \der_3\eta_1
        \der_3\eta_2$$ $$ -
  2 \der_1\eta_1 \der_1\eta_3 \der_2\eta_1 \der_2\eta_2 \der_3\eta_1
        \der_3\eta_2 +
  4  \der_1\eta_1 \der_1\eta_3 \der_2\eta_1 \der_3\eta_1 \der_3\eta_2
        \der_3\eta_3$$ $$ -
  2 \der_1\eta_2 \der_2\eta_1 \der_2\eta_2 \der_2\eta_3 \der_3\eta_1
        \der_3\eta_2 -
  \der_1\eta_2 \der_2\eta_1 \der_2\eta_3 \der_3\eta_1 \der_3\eta_2
      \der_3\eta_3$$ $$ -
  \der_1\eta_3 \der_2\eta_1 \der_2\eta_2 \der_3\eta_1 \der_3\eta_2
\der_3\eta_3,$$

$$H_{\eta_2\eta_3}=$$
$$-3\der_1\eta_1\der_2\eta_1\der_2\eta_2\der_2\eta_3\der_3\eta_1
\der_3\eta_2 - 3 \der_1\eta_1 \der_2\eta_1 \der_2\eta_3 \der_3\eta_1
\der_3\eta_2 \der_3\eta_3.$$

Thus, $\eta_1\eta_2\eta_3\der_1^2$-part of $F\circ G$ by Lemma \ref{Altynai}
is equal to $\eta_1\eta_2\eta_3\der_1^2$-part of
$$F\circ
(G_{\eta_1\eta_2;\der_1^2}+G_{\eta_1\eta_3;\der_1^2}+G_{\eta_2\eta_3;\der_1^2}).$$
So, $\eta_1\eta_2\eta_3\der_1^2$-part of $F\circ G$ is equal to
$$\eta_1\eta_2\eta_3(
\der_1(D)\circ H_{\eta_2\eta_3;\der_1^2}+
\der_2(D)\circ H_{\eta_1\eta_3;\der_1^2}+
\der_3(D)\circ H_{\eta_1\eta_2;\der_1^2})\der_1^2$$
Calculations show that this expression is equal to $-G_1.$
So, we obtain that $\eta_1\eta_2\eta_3\der_1^2$-part of $F^{\circ 2}\bullet
F^{\circ 3}+F\circ(F\bullet F^{\circ 3})$ is equal 0.

Similar calculations show that sums of
$\eta_i\eta_j\der_1^2$-parts of $F^{\circ 2}\bullet F^{\circ 3}$ and $F\circ
(F\bullet F^{\circ 3})$ are
also vanish, if $i\ne j.$ So, we have established that
$$F^{\circ 2}\bullet F^{\circ 3}+F\circ(F\bullet F^{\circ 3})=0.$$

\section{Qubic differential part of $D^{10}$}
In this section we use denotions and results of calculations
of section \ref{quadraticpart}.

\begin{lm}\label{x}
$F^{\circ 3}\bullet F^{\bullet 2}=0$
\end{lm}

{\bf Proof.}
Recall that  $G=F\bullet F^{\circ 3}.$ By associativity and
super-commutativity of bullet-multiplication (proposition \ref{???}) we have
$$F^{\circ 3}\bullet(F\bullet F)=F\bullet(F\bullet F^{\circ 3}).$$
So, $\eta_1\eta_2\eta_3\der_1^2$-part of $F^{\circ 3}\bullet F^{\bullet 2}$
is equal to
$$F\bullet (\eta_1\eta_2 H_{\eta_1\eta_2}+\eta_1\eta_3
H_{\eta_1\eta_3}+\eta_2\eta_3H_{\eta_2\eta_3})\der_1^2$$
$$=\eta_1\eta_2\eta_3(\der_1\eta_1\der_1\bullet
H_{\eta_2\eta_3}\der_1^2-\der_2\eta_1 \der_1\bullet
H_{\eta_1\eta_3}\der_1^2+\der_3\eta_1\der_1\bullet
H_{\eta_1\eta_2}\der_1^2).$$
By results of section \ref{quadraticpart} it is easy to obtain that
$$\der_1\eta_1H_{\eta_2\eta_3}= 0,$$
$$\der_2\eta_1 H_{\eta_1\eta_3}=0,$$
$$\der_3\eta_1 H_{\eta_1\eta_2}=0.$$
So,
$\eta_1\eta_2\eta_3\der_1^3$-part of $F^{\circ 3}\bullet F^{\bullet 2}$ is
$0.$ Since number of bullets is two, $\eta_i\eta_j\der_1^3$-parts and
$\eta_i\der_1^3$-parts of $F^{\circ 3}\bullet F^{\bullet 2}$ are also $0.$
So, $\der_1^3$-part of
$F^{\circ 3}\bullet F^{\bullet 2}$ vanishes. By symmetry
$\der^{\alpha}$-part of  $F^{\circ 3}\bullet F^{\bullet 2}$ vanishes also
for any $\alpha\in\Gamma_3,$ such that $|\alpha|=3.$
Lemma is proved.

\begin{lm}
$F\bullet F^{\circ 2}\bullet F^{\circ 2}=0.$
\end{lm}

{\bf Proof.} Let $R=F^{\circ 2}\bullet F^{\circ 2}.$ We use calculations
on $F^{\circ 2}$ ( section \ref{secondlsym}) to obtain
$$R_{\eta_1\eta_2;\der_1^2}$$
$$=\eta_1\eta_2
(8\der_1\eta_1 \der_2\eta_1 \der_3\eta_1 \der_1\eta_2 \der_2\eta_2
        \der_2\eta_3 -
  4  \der_1\eta_1 \der_2\eta_1 \der_3\eta_1 \der_1\eta_2 \der_2\eta_3
        \der_3\eta_3 $$ $$+
  2  \der_1\eta_1 \der_2\eta_1 \der_3\eta_1 \der_2\eta_2 \der_1\eta_3
        \der_3\eta_3+
  6  \der_1\eta_1 \der_2\eta_1 \der_3\eta_1 \der_3\eta_2 \der_1\eta_3
        \der_2\eta_3$$ $$ -
  2  \der_2\eta_1 \der_3\eta_1 \der_1\eta_2 \der_2\eta_2 \der_2\eta_3
        \der_3\eta_3 +
  2  \der_2\eta_1 \der_3\eta_1 \der_2\eta_2 \der_3\eta_2 \der_1\eta_3
        \der_2\eta_3 + $$ $$
  4  \der_2\eta_1 \der_3\eta_1 \der_3\eta_2 \der_1\eta_3 \der_2\eta_3
        \der_3\eta_3)\der_1^2,$$

$$R_{\eta_1\eta_3;\der_1^2}$$
$$=\eta_1\eta_3(
-2  \der_1\eta_1 \der_2\eta_1 \der_3\eta_1 \der_1\eta_2 \der_2\eta_2
        \der_3\eta_3 -
  6 \der_1\eta_1 \der_2\eta_1 \der_3\eta_1 \der_1\eta_2 \der_3\eta_2
        \der_2\eta_3 $$ $$-
  4 \der_1\eta_1 \der_2\eta_1 \der_3\eta_1 \der_2\eta_2 \der_3\eta_2
        \der_1\eta_3+
  8 \der_1\eta_1 \der_2\eta_1 \der_3\eta_1 \der_3\eta_2 \der_1\eta_3
        \der_3\eta_3$$ $$ -
  4 \der_2\eta_1 \der_3\eta_1 \der_1\eta_2 \der_2\eta_2 \der_3\eta_2
        \der_2\eta_3 +
  2 \der_2\eta_1 \der_3\eta_1 \der_1\eta_2 \der_3\eta_2 \der_2\eta_3
        \der_3\eta_3$$ $$ -
  2 \der_2\eta_1 \der_3\eta_1 \der_2\eta_2 \der_3\eta_2 \der_1\eta_3
        \der_3\eta_3)\der_1^2,$$

$$R_{\eta_2\eta_3;\der_1^2}$$
$$=\eta_2\eta_3(
-6  \der_1\eta_1 \der_2\eta_1 \der_3\eta_1 \der_2\eta_2 \der_3\eta_2
        \der_2\eta_3 +
  6 \der_1\eta_1 \der_2\eta_1 \der_3\eta_1 \der_3\eta_2 \der_2\eta_3
        \der_3\eta_3)\der_1^2.$$
We have
$$\eta_1\der_1\eta_1\der_1\bullet R_{\eta_2\eta_3;\der_1^2}=0,
$$
$$
\eta_2\der_2\eta_1\der_1\bullet R_{\eta_1\eta_3;\der_1^2}=0,
$$
$$
\eta_3\der_3\eta_1\der_1\bullet R_{\eta_1\eta_2;\der_1^2}=0.
$$
Therefore  $\eta_1\eta_2\eta_3\der_1^3$-part of
$F\bullet F^{\circ 2}\bullet F^{\circ 2}$ is equal to
$$\eta_1\der_1\eta_1\der_1\bullet R_{\eta_2\eta_3;\der_1^2}
+\eta_2\der_2\eta_1\der_1\bullet R_{\eta_1\eta_3;\der_1^2}
+\eta_3\der_3\eta_1\der_1\bullet R_{\eta_1\eta_2;\der_1^2}$$
$$=0.$$
By symmetry, $\eta_1\eta_2\eta_3\der_1^{\alpha}$-part of
$F\bullet F^{\circ 2}\bullet F^{\circ 2}$ are also $0$ for any $\alpha\in
\Gamma_3,$ such that $|\alpha|=3.$
As we mentioned above $\eta_i$- and $\eta_i\eta_j$-parts  of elements
obtained by two bullets are equal to $0.$ Lemma is proved.


By Lemma \ref{mainlemma}
$$\tau_3(F^5)=
5(2 F^{\circ 3}\bullet F^{\bullet 2}+3F\bullet F^{\circ 2}\bullet F^{\circ
2}).$$
Therefore, we come to the following

{\bf Conclusion.} $\tau_3(D^{10})=0.$

{\bf Proof of Theorem \ref{main1}}

By Theorem 0.1 of \cite{odd}
if $\eta_{i_1}\cdots \eta_{i_r} \der^{\alpha}$-part of
$D^{10}$ is nonzero, then $$|\alpha|\le 3.$$
So,
$$D^{10}=\tau_1(D^{10}).$$

\section{$N$-commutators and super-derivations}

In this section we explain how escort invariants appear in calculating
powers of odd derivations.

Suppose now $I=\{1,\ldots,n\}$ and $D=\sum_{i=1}^nu_i\der_i\in Der\,\LL$
odd super-derivation. For $\alpha\in{\bf Z}^n_+$ set $$x^{(\alpha)}=
\frac{x^{\alpha}}{\alpha!}.$$
Denote by $Supp(s_k)$
a set of $k$-typles $\{(\alpha^{(1)},i_1),\cdots,(\alpha^{(k)},i_k)\}$
with  $\alpha^{(1)},\ldots,\alpha^{(k)}\in {\bf Z}^n_+, i_1,\ldots,
i_k\in I,$
such that $\sum_{p=1}^k\alpha^{(p)}-\epsilon_{i_p}$ has a form $-\beta$ for
some  $0\ne \beta\in {\bf Z}^n_+.$

\begin{th} \label{baisary}
$$k!D^k=$$

$$
\sum
\der^{\alpha}(u_{i_1})\der^{\beta}(u_{i_2})\cdots \der^{\gamma}(u_{i_k})
esc(s_k)(x^{(\alpha)}\der_{i_1},x^{(\beta)}\der_{i_2},\ldots,x^{(\gamma)}
\der_{i_k}),
$$
where summation is given by
$\{(\alpha,i_1),(\beta,i_2),\ldots,(\gamma,i_k)\}\in Supp(s_k).$
\end{th}

If we use order on basic elements $x^{\alpha}\der_i$ we can omit the coefficient $k!:$
$$D^k=
\sum_{(\alpha^{(1)},i_1)<
\cdots <(\alpha^{(k)},i_k)}
\der^{\alpha^{(1)}}(u_{i_1})\cdots \der^{\alpha^{(k)}}(u_{i_k})
esc(s_k)(x^{(\alpha^{(1)})}\der_{i_1},\ldots,x^{(\alpha^{(k)})}\der_{i_k}).
$$

{\bf Proof.} Recall that $U={\bf C}[x_1,\ldots,x_n]$ and $\der_i$ are partial
derivations of $U.$
Let $Gr_k$ be a Grassman
algebra with exterior generators $\eta_1,\ldots,\eta_k,$ i.e., it is associative super-commutative algebra of dimension $2^k.$ For
$U={\bf C}[x_1,\ldots,x_n]$ take its Grassman envelope
$${\cal U}= U\otimes Gr_k.$$
Prolong derivation $\der_i\in Der\,U$ to a derivation of ${\cal U}$ by
$$\der_i(v\otimes \omega)=\der_i(v)\otimes \omega.$$
We obtain commuting system of even derivations
${\cal D}=\{\der_1,\ldots,\der_n\}$
of ${\cal U}.$ For any $f_1,\ldots,f_n\in {\cal U}$ and
$\alpha\in {\bf Z}^n_+$ elements $\der^{\alpha}u_j$ are odd.
So, we obtain ${\cal D}$-differential super-algebra ${\cal U}$ and we can
consider its algebra of super-derivations
 ${\cal L}=<f\der_i: f\in {\cal U}> $
and its algebra of super-differential operators
$${\cal D}if\!f=<f\der^{\alpha}: \alpha\in {\bf Z}^n_+, f\in {\cal U}>.$$
We can endow ${\cal D}if\!f$ by composition operation, by left-symmetric multiplication and
by bullet multiplication. In particular, we can consider
${\cal L}$ as a  left-symmetric algebra and as a  super-Lie algebra. Thus,
$${\cal L}\cong W_n\otimes Gr_k$$
is isomorphic to a current algebra with coefficients not in Laurent
polynomials as usual, but in exterior algebra.

We see that for any $f_1,\ldots,f_n$ we can consider a homomorphism
$$\LL_n\rightarrow {\cal U}, u_i\mapsto f_i, i=1,\ldots,n.$$
and this homomorphism can be extended to a homomorphism of left-symmetric
or Lie algebras
$$Der\,\LL_n\rightarrow {\cal L}$$
and to a homomorphism of associative (left-symmetric) algebras
$$Dif\!f\rightarrow {\cal D}if\!f$$
We can use this homomorphism in calculating $F^k$
for $F=\sum_{i=1}^nf_i\der_i\in {\cal L}.$ In other words,
in the formula for $D^k$  we can make substitutions $u_i\mapsto f_i$ and
calculate obtained expressions in $\cal U.$

Use this method for calculating coefficients $\lambda_{\{(\alpha,i_1),(\beta,i_2),\ldots,(\gamma,i_k);\mu\}},$ where
$$D^k=\sum
\lambda_{\{(\alpha,i_1),(\beta,i_2),\ldots,(\gamma,i_k);\mu\}}
\der^{\alpha}(u_{i_1})\der^{\beta}(u_{i_2})\cdots
\der^{\gamma}(u_{i_k})\der^{\mu}.$$
Since numbers of $u_i$-indexes and $\der_i$-indexes are equal, summation here
is done by  $\alpha,\beta, \ldots,\gamma,\mu\in {\bf Z}^n_+$
and $i_1,\ldots,i_k\in\{1,\ldots,n\}$ such that
$\alpha+\beta+\cdots+\gamma+\mu=\sum_{s=1}^k\epsilon_{i_s}.$
In other words, summation here is done by
$\{(\alpha,i_1),(\beta,i_2),\ldots,(\gamma,i_k)\}\in
Supp_{|\mu|}(s_k).$

Take
$$F=X_1\otimes \eta_1+\cdots+X_k\otimes \eta_k\in {\cal L},$$
where $X_i\in W_n, i=1,\ldots,k$ are even elements.
It is evident that
$$F^k=s_k(X_1,\ldots,X_k)\otimes (\eta_1\cdots \eta_k).$$
On the other hand, if $X_1=x^{\alpha^{(1)}}\der_{i_1}, X_2=x^{\alpha^{(2)}}\der_{i_2},\ldots,X_k=x^{\alpha^{(k)}}\der_{i_k},$ then $F$ can be presented
in the form
$$F=\sum_{i=1}^kf_i\der_i\in {\cal L},$$
where
$$f_i= \sum_{s:  i_s=i}x^{\alpha^{(s)}} \otimes \eta_s\in {\cal U}$$
summation by $s$ such that $i_s=i.$ So, substitutions
$$u_i\mapsto \sum_{s:  i_s=i}x^{\alpha^{(s)}} \otimes \eta_s\in {\cal U}$$
in $D^k$ and making calculations in ${\cal U}$ gives us on the one side
$$\lambda_{\{\alpha^{(1)},i_1),\ldots,(\alpha^{(k)},i_k);\mu\}}
\alpha^{(1)}!\cdots \alpha^{(k)}!\der^{\mu}\otimes \eta_1\cdots \eta_k+ Y,$$
where
$$Y\in <x^{\alpha}\der^{\beta}\otimes Gr_k : |\alpha|>0, \alpha,\beta\in {\bf Z}^n_+>,$$
and one the other side
$$s_k(x^{\alpha^{(1)}}\der_{i_1},\ldots,x^{\alpha^{(k)}}\der_{i_k})
\otimes  \eta_1\cdots \eta_k.$$
Take projections ${\cal D}if\!f\rightarrow <1>\otimes \eta_1\ldots \eta_k$
from the both parts. We have
$$\lambda_{\{\alpha^{(1)},i_1),\ldots,(\alpha^{(k)},i_k);\mu\}}
\alpha^{(1)}!\cdots \alpha^{(k)}!\der^{\mu}= esc(s_k)
(x^{\alpha^{(1)}}\der_{i_1},\ldots,x^{\alpha^{(k)}}\der_{i_k}).$$
Thus,
$$esc(s_k)
(x^{(\alpha^{(1)})}\der_{i_1},\ldots,x^{(\alpha^{(k)})}\der_{i_k})=
\lambda_{\{\alpha^{(1)},i_1),\ldots,(\alpha^{(k)},i_k);\mu\}}
\der^{\mu}
$$
that we need to prove. $\square$

Recall that $k$-commutator $s_k$ is called well defined on $W_n$ if
$$\forall X_1,\ldots,X_k\in W_n\Rightarrow s_k(X_1,\ldots,X_k)\in W_n$$
Denote by $s_k^{\circ}$ a map $\wedge^kW_n\rightarrow W_n$ given by
$$s_k^{\circ}(X_1,\ldots,X_k)=\sum_{\sigma\in Sym_k}sign\,\sigma\,
X_{\sigma(1)}\circ (X_{\sigma(2)}\circ (\cdots (X_{\sigma(k-1)}\circ
X_{\sigma(k)}))),$$
where  $W_n$ is considered as a left-symmetric algebra under multiplication
$f\der_i\circ g\der_j=f\der_i(g)\der_j.$

\begin{crl} The following conditions are equivalent:
\begin{itemize}
\item $D^k\in Der\,\LL$
\item $D^k=D^{\circ k}$
\item $s_k$ is well defined operation on $W_n.$
\item $s_k=s_k^{\circ}.$
\end{itemize}
\end{crl}

Theorem \ref{baisary} has two-fold applications. We use it in constructing
$D^k$ by $s_k$ and vice versa one can use $D^k$ in calculating
$k$-commutators.

{\bf Example 1.} Let  $k=1.$ Then
$s_1(x^{\alpha}\der_i)=x^{\alpha}\der_i,$ therefore
$supp(s_1)=<\der_i: i=1,\ldots,n>$ and
$Supp(s_1)=\{(0,1),\ldots,(0,n)\}.$ Hence
$$D^1=\sum_{i=1}^n \der^0(u_i)s_1(\der_i)=\sum_{i=1}^n u_i\der_i,$$
that we know well.

{\bf Example 2.} Let us calculate
$Y=s_{11}(X_1,\ldots,X_{11})$ for
$X_i=\der_i, 1\le i\le 3,$ $X_4=x_1\der_1-x_3\der_3,$ $X_5=x_2\der_1,$
$X_6=x_3\der_1,$ $X_7=x_1\partial_2,$ $X_8=x_2\partial_2
-x_3\partial_3,$ $X_9=x_1\partial_3,$
$X_{10}=x_2\partial_3,$
$X_{11}=x_3^2\partial_1.$ Let $Gr_{11}$ be Grassman algebra generated by $11$
odd elements $\eta_1,\ldots,\eta_{11}.$
Take ${\cal U}={\bf C}[x_1,x_2,x_3]\otimes Gr_{11}.$
Recall that $x_i$ and $\der_i$ are even variables.
Consider a homomorphism of super-differential polynomials
algebra $\LL_3$ to $\cal U$
given by
$$u_1\mapsto \eta_1+x_1 \eta_4 +x_2\eta_5+x_3\eta_6+x_3^2\eta_{11},$$
$$u_2\mapsto \eta_2+x_1\eta_7+x_2\eta_8,$$
$$u_3\mapsto \eta_3-x_3\eta_4-x_3\eta_8+x_1\eta_9+x_2\eta_{10}.$$
In other words make in the formulas for $\tau_1(D^{11})$ and $\tau_2(D^{11})$
 corresponding substitutions.
Make all calculations in $\LL_3$ using the formula
$\der_i(v\otimes \eta)=\der_i(v)\otimes\eta,$ $v\in {\bf C}[x_1,x_2,x_3],
\eta\in Gr_{11}.$
One obtains that the linear part of $Y$ is equal to $0$ and
the quadratic part of $Y$ is equal to $80\partial_1^2.$ So,
$Y=80\der_1^2.$

The following results about $N$-commutators on
$Vect(2)$ and $Vect_0(2)$ was established in \cite{N-commutators}.
$6$-commutator on $W_2$ is well defined and it has one escort
invariant
$$escort_{231}:L_0\otimes L_1\rightarrow R(\pi_1)\cong L_{-1},$$
$$escort_{231}(a,X)=d(a\circ Div\,X)+Div\,a\; d\,Div\,X-3 d\,Div(a\circ X).$$
$s_6=0$ is identity on $S_1$ and $5$-commutator is well defined on
$S_1$ and it has one escort invariant
$$escort_{221}:R(2\pi_1)\otimes R(3\pi_1)\rightarrow R(\pi_1)\cong L_{-1},$$
$$escort_{221}(a,X)=-3 d\,Div(a\circ X).$$

Below we do similar things for $Vect(3)$ and $Vect_0(3)$. Since
calculations are too tedious and similar to calculations given above we
formulate final results and give some examples.

\section{$13$-commutator on $Vect(3)$}

\begin{th} \label{27march} If $n=3,$ then  $13$-commutator on $W_3$
$$X_1,\ldots,X_{13}\in W_3\Rightarrow s_{13}(X_1,\ldots,X_{13})\in W_3.
$$
It has one escort invariant
$$escort_{382}=escort(s_{13}): L_0\otimes \wedge^2 L_1\rightarrow \wedge^2 R(\pi_1)\cong
L_{-1}$$
defined by
$$escort_{382}(a,X,Y)=$$

$$-d(a\circ Div\,X)\wedge d(Div\,Y)+d(a\circ Div\,Y)\wedge d(Div\,X)$$
$$-2 (Div\,a) d(Div\,X)\wedge d(Div\,Y)$$
$$+4\,d(Div\,X\circ a)\wedge d(Div\,Y)+4\, d(Div\,X)\wedge d(Div\,Y\circ a)$$
$$+8 (da\stackrel{\circ}{\wedge} dX)\circ Div\,Y-
8(da\stackrel{\circ}{\wedge} dY)\circ Div\,X.$$

%
\end{th}

\begin{crl}
\label{kulsat}
 $escort_{382}$ induces a homomorphism
$\bar L_0\otimes \bar L_1\otimes R(2\pi_1) \rightarrow \wedge^3 R(\pi_1)$
by
$$(a\der_i,v\der_j,w)\mapsto da\wedge d\der_i(v)\wedge d\der_j(w).$$
\end{crl}

\begin{crl} $s_{13}=0$ is identity on $Vect_0(3).$
Moreover $s_{12}=0$ is identity on $Vect_0(3).$
\end{crl}

Let $$G_{ij}(a)=\der_i(a)\der_j-\der_j(a)\der_i,$$
$$\tilde u=u(x_1\der_1+x_2\der_2+x_3\der_3).$$
Take place isomorphisms of $sl_n$-modules
$$
L_1=\bar L_1+\tilde L_1,$$
where
$$\bar L_1=<X : Div(X)=0>\cong R(2\pi_1+\pi_{n-1})\cong R(2\pi_1+\pi_{n-1}),$$
$$\tilde L_1\cong R(\pi_1).$$
Then $\bar L_1$ is generated by elements of the form
$G_{ij}(x^{\alpha}),$ where $i<j$ $\alpha\in {\bf Z}^n_+, |\alpha|=3$
and $\tilde L_1$ is has a basis $\{\tilde x_i: i=1,\ldots,n\|.$

Let us give construction of escort invariant in terms of ${\bar L}_i$
and ${\tilde L}_i.$ We see that
$escort(s_{13})(a,X,Y)=0,$ if $X,Y\in \tilde L_1$ or $X,Y\in \bar L_1$
or $a=\tilde 1=x_1\der_1+x_2\der_2+x_3\der_3.$
Below we use the following notation
$$a^{(i,j,k)}=\der_1^i\der_2^j\der_3^s(a).$$
Non-zero components of $escort(s_{13})(a,X,Y)$ can be given by:
$$escort(s_{13})(G_{12}(a),G_{12}(b),\tilde x_1)=$$

$$(-32a^{(1,0,1)}b^{(0,3,0)}+ 32a^{(1,1,0)} b^{(0,2,1)}
-32 a^{(0,2,0)}b^{(1,1,1)}+32a^{(0,1,1)}b^{(1,2,0)})\der_1 $$
$$+(32a^{(1,0,1)}b^{(1,2,0)}
-16 a^{(2,0,0)} b^{(0,2,1)}+16a^{(0,2,0)}b^{(2,0,1)}
- 32a^{(0,1,1)}b^{(2,1,0)})\der_2$$
$$+(-32a^{(1,1,0)}b^{(1,2,0)} +16a^{(2,0,0)} b^{(0,3,0)}
+16a^{(0,2,0)}b^{(2,1,0)})\der_3,$$

$$escort(s_{13})(G_{12}(a),G_{12}(b),\tilde x_2)=$$

$$(32a^{(1,0,1)}b^{(1,2,0)}
-  16a^{(2,0,0)} b^{(0,2,1)}
+ 16a^{(0,2,0)}b^{(2,0,1)}
- 32a^{(0,1,1)}b^{(2,1,0)})\der_1
$$
$$+
(32 a^{(2,0,0)} b^{(1,1,1)}
-   32a^{(1,1,0)}b^{(2,0,1)}
- 32a^{(1,0,1)}   b^{(2,1,0)}
+  32a^{(0,1,1)}   b^{(3,0,0)} )\der_y
$$
$$+(-   16a^{(2,0,0)}b^{(1,2,0)}
+   32a^{(1,1,0)}b^{(2,1,0)}
-   16a^{(0,2,0)}b^{(3,0,0)})\der_3
$$

$$escort(s_{13})(G_{12}(a),G_{12}(b),\tilde x_3)=$$

$$(
-32a^{(1,1,0)}b^{(1,2,0)}
+16a^{(2,0,0)} b^{(0,3,0)}
+16a^{(0,2,0)}b^{(2,1,0)} )\der_1
$$
$$+(
- 16a^{(2,0,0)}b^{(1,2,0)}
+ 32a^{(1,1,0)}b^{(2,1,0)}
- 16a^{(0,2,0)}b^{(3,0,0)}
)\der_2
$$

$$escort(s_{13})(G_{12}(a),G_{23}(b),\tilde x_1)=$$

$$(
-16a^{(1,0,1)} b^{(0,2,1)}
- 16a^{(0,2,0)} b^{(1,0,2)}
+ 16a^{(1,1,0)} b^{(0,1,2)}
+ 16a^{(0,1,1)} b^{(1,1,1)}
)\der_2
$$
$$+(
+ 16a^{(0,2,0)}b^{(1,1,1)}
- 16a^{(0,1,1)}b^{(1,2,0)}
- 16a^{(1,1,0)}b^{(0,2,1)}
+ 16a^{(1,0,1)}b^{(0,3,0)}
)\der_3
$$

$$escort(s_{13})(G_{12}(a),G_{23}(b),\tilde x_2)=$$

$$(
-16   a^{(1,0,1)}b^{(0,2,1)} -
  16a^{(0,2,0)}   b^{(1,0,2)} +
  16   a^{(1,1,0)} b^{(0,1,2)}
+   16a^{(0,1,1)}   b^{(1,1,1)} )\der_1
$$
$$+(32a^{(1,1,0)} b^{(1,0,2)}
+ 32a^{(1,0,1)} b^{(1,1,1)}
- 32 a^{(2,0,0)} b^{(0,1,2)}
- 32a^{(0,1,1)}b^{(2,0,1)}
)\der_2$$
$$
+(- 48a^{(1,1,0)}b^{(1,1,1)} -
  16a^{(1,0,1)}b^{(1,2,0)}
+ 32a^{(2,0,0)}b^{(0,2,1)}$$
$$
+ 16a^{(0,2,0)}b^{(2,0,1)} +
  16a^{(0,1,1)}b^{(2,1,0)})\der_3
$$

$$escort(s_{13})(G_{12}(a),G_{23}(b),\tilde x_3)=$$

$$(
16   a^{(1,0,1)}b^{(0,3,0)} -
  16   a^{(1,1,0)}b^{(0,2,1)} +
  16a^{(0,2,0)}   b^{(1,1,1)}
- 16a^{(0,1,1)}   b^{(1,2,0)} )\der_1$$
$$+(- 48a^{(1,1,0)}b^{(1,1,1)}
-   16a^{(1,0,1)}b^{(1,2,0)}
+   32a^{(2,0,0)} b^{(0,2,1)}$$
$$
+   16a^{(0,2,0)}b^{(2,0,1)} +
  16a^{(0,1,1)}b^{(2,1,0)}
)\der_2
$$
$$+(-  32a^{(2,0,0)} b^{(0,3,0)}
+   64a^{(1,1,0)}b^{(1,2,0)}
-   32a^{(0,2,0)}b^{(2,1,0)})\der_3.$$
\medskip

\noindent If $\{i,j,s,k\}\subseteq \{1,2,3\},$ then any two pairs
$(i,j),(s,k)$ has at least one common element. Therefore
by symmetry one can easily write other formulas for
$escort(s_{13})(G_{ij}(a),G_{sk}(b),\tilde x_r).$

Let us show how to use theorem \ref{27march} in calculation
of $13$-commutator on $Vect(3).$

{\bf Example 1.} Take
$$X_1=\der_1, X_2=\der_2, X_3=\der_3,
X_4=x_1\der_1, X_5=x_2\der_1, X_6=x_3\der_1, X_7=x_1\der_2,$$
$$X_8=x_2\der_2,
X_9=x_3\der_2, X_{10}=x_1\der_3, X_{11}=x_2\der_3,X_{12}=x_1 x_2\der_3,
X_{13}=x_3^2\der_3.$$
We see that the number of elements of grade $-1$ is $3$ and the  number
elements of grade $0$ is $8$. In $0$-part here appear all base elements of
$gl_3$ except $a=x_3\der_3.$ Elements of the grade $1$ are two: $X=X_{12}$ and
$Y=X_{13}.$ So, to calculate $13$-commutator of $13$ elements $X_1,\ldots,
X_{13},$ we denote it $s_{13}(X_1,\ldots,X_{13}),$ we need to calculate
$escort(s_{13})(a, X,Y).$ We have
$$Div\,X=\der_3(x_1x_2)=0,$$
therefore,
$$-d(a\circ Div\,X)\wedge d(Div\,Y)+d(a\circ Div\,Y)\wedge d(Div\,X)
-2 (Div\,a) d(Div\,X)\wedge d(Div\,Y)=0.$$
Further,
$$Div\,a=\der_3(x_3)=1,\quad Div\,X\circ a=Div(x_1x_2\der_3(x_3)\der_3)=0,
$$
and
$$+4\,d(Div\,X\circ a)\wedge d(Div\,Y)+
4\, d(Div\,X)\wedge d(Div\,Y\circ a)=0.$$
Finally, $Div \,Y=2 x_3$ and
$$8\,\{\sum_{i,j=1}^3
\der_i(Div\,Y) d(a(x_j))\wedge d(\der_j(X(x_i)))
-
\der_i(Div\,X) d(a(x_j))\wedge d(\der_j(Y(x_i)))\}=
$$
$$8 \sum_{i,j=1}^3 \der_i(2x_3)d(a(x_j))\wedge d(\der_j(x_1x_2\der_3(x_i)))=$$
$$16
\sum_{j=1}^3 d(a(x_j))\wedge d(\der_j(x_1x_2))=0.$$
Therefore,
$$escort(s_{13})(x_3\der_3, x_1x_2\der_3,x_3^2\der_3)=0.$$

{\bf Example 2.}
Now change in example 1 $X_{12}$ to
$$X_{12}=x_1x_3\der_3,$$
other elements are as before. Two vector fields, nameley $X=X_{12}$ and
$Y=X_{13}$ have non-constant divergences:
$$Div\,X=x_1, Div\,Y=2x_3.$$
So, we can expect that $s_{11}(X_1,\ldots,X_{13})$ might be non-trivial
vector field. We have
$$-d(a\circ Div\,X)\wedge d(Div\,Y)+d(a\circ Div\,Y)\wedge d(Div\,X)=$$
$$2 dx_3\wedge dx_1,$$

$$-2 (Div\,a) d(Div\,X)\wedge d(Div\,Y)=-4 dx_1\wedge dx_3,$$

$$4\,d(Div\,X\circ a)\wedge d(Div\,Y)+
4\, d(Div\,X)\wedge d(Div\,Y\circ a)=$$
$$8\,dx_1\wedge dx_3+8dx_1\wedge dx_3,$$

$$8\,\{\sum_{i,j=1}^3
\der_i(Div\,Y) d(a(x_j))\wedge d(\der_j(X(x_i)))-
\der_i(Div\,X) d(a(x_j))\wedge d(\der_j(Y(x_i)))\}=
$$
$$8\,\{2 dx_3\wedge d(\der_3(x_1x_3))=$$
$$-16\,dx_1\wedge dx_3.$$
Thus,
$$escort(s_{13})(a,X,Y)=-6 dx_1\wedge dx_3.$$
Since the isomorphism
$\wedge^2 R(\pi_1)\cong R(\pi_2)$ is established by
$$dx_1\wedge dx_2\mapsto \der_3, dx_1\wedge dx_3\mapsto -\der_2,
dx_2\wedge dx_3\mapsto \der_1$$
this means that
$$s_{13}(X_1,\ldots,X_{13})=
esc(s_{13})(X_1,\ldots,X_{13})=
6\der_2.$$

{\bf Example 3.}
Let now
$$X_{12}=x_1 x_3\der_3, X_{13}=x_2x_3\der_3.$$
other elements as above. Then
$$-d(a\circ Div\,X)\wedge d(Div\,Y)+d(a\circ Div\,Y)\wedge d(Div\,X)=0,$$

$$-2 (Div\,a) d(Div\,X)\wedge d(Div\,Y)=-2 dx_1\wedge dx_2,$$

$$4\,d(Div\,X\circ a)\wedge d(Div\,Y)+
4\, d(Div\,X)\wedge d(Div\,Y\circ a)=$$
$$4\,dx_1\wedge dx_2+4\,dx_1\wedge dx_2=8\,dx_1\wedge dx_2,$$

$$\sum_{i,j=1}^3
\der_i(Div\,Y) d(a(x_j))\wedge d(\der_j(X(x_i)))-
\der_i(Div\,X) d(a(x_j))\wedge d(\der_j(Y(x_i)))=0.
$$
Thus,
$$escort(s_{13})(a,X,Y)=6 dx_1\wedge dx_2,$$
and
$$s_{13}(X_1,\ldots,X_{13})=escort(s_{13})(X_1,\ldots,X_{13})=
6\der_3.$$

{\bf Example 4.} Let now all
$Y_i=X_i$ as before if $i<12$ and
$$Y_{12}=x_1x_3\der_3, Y_{13}=x_2x_3^2\der_3.$$
Then
$$s_{13}(Y_1,\ldots,Y_{13})=E_{x_3^2\der_3}(Y_{13})
esc(s_{13})(X_1,\ldots,X_{11},x_1x_3\der_3,x_3^2\der_3)+$$
$$
E_{x_2x_3\der_3}(Y_{13})esc(s_{13})(X_1,\ldots,X_{11},x_1x_3\der_3,
x_2x_3\der_3)=$$
(by results of example 2 and 3 )
$$ 6 x_2\der_2 + 12 x_3\der_3.$$


\section{ $10$-commutator on $Vect(3)$ and $Vect_0(3)$}

Recall some denotions:
$$U_k=\left< x^{\alpha} | |\alpha|=k\right >,$$
the multiplication $\circ$ is left-symmetric and
$\stackrel{\circ}{\wedge}$ means
wedge-product corresponding to left-symmetric multiplication:
$$L_1\times U_2\rightarrow \wedge^2U_1,\qquad (u\der_i,v)\mapsto du\der_i\stackrel{\circ}{\wedge} dv:=du\wedge
d\der_i(v).$$  Below expressions like
$Div\, a\circ X$ will mean $Div\,(a\circ X).$

\begin{th} \label{3april2003}$10$-commutator is well defined on $L=Vect(3)$:
$$X_1,\ldots,X_{10}\in L\Rightarrow s_{10}(X_1,\ldots,X_{10})\in L.
$$
It has three escort invariants
$$escort_{271}: R(\pi_1)\otimes \wedge^2 L_0\otimes L_2
\rightarrow \wedge^2 R(\pi_1)\cong
L_{-1},$$

$$escort_{3601}:\wedge^3L_0\otimes L_2\rightarrow \wedge R(\pi_1)\cong L_{-1},$$and
$$escort_{352}:\wedge^4L_0\otimes\wedge^2L_1\rightarrow \wedge^2R(\pi_1)\cong L_{-1}.$$
They can be given by
$$escort_{271}(u,a,b,X)=$$

$$du\wedge (11 d(Div\;([a,b]\circ X))+21 d(a\circ Div(b\circ X)-
b\circ Div(a\circ X))-44 d([a,b]\circ Div\;X))$$
$$-32 (d(a\circ u)\wedge d(Div(b\circ X))-d(b\circ u)\wedge d(Div(a\circ X)))$$
$$-50 (d(a\circ u)\wedge d(b\circ Div\,X)-d(b\circ u)\wedge d(a\circ Div\,X))$$
$$+Div\,a\; du\wedge d(2 b\circ Div\,X+9 Div(b\circ X))-
Div\,b\; du\wedge d(2 a\circ Div\,X+9 Div(a\circ X))$$
$$+8(Div\,a\;db\stackrel{\circ}{\wedge}d(u\,Div\,X)-
Div\,b\; da\stackrel{\circ}{\wedge}d(u\,Div\,X))$$
$$+12 (da\stackrel{\circ}{\wedge}d((b\circ X)\circ u)-db\stackrel{\circ}{\wedge}
d((a\circ X)\circ u))$$
$$-28 d[a,b]\stackrel{\circ}{\wedge} d(X\circ u)$$
$$+16 (da\stackrel{\circ}{\wedge}d (X\circ(b\circ u))- db\stackrel{\circ}{\wedge} d(X\circ (a\circ u))),$$

\bigskip

$$escort_{3601}(a,b,c,X)=$$
$$-6(da\stackrel{\circ}{\wedge} d(b\circ(Div\;(c\circ  X)))
+db\stackrel{\circ}{\wedge} d(c\circ(Div\;(a\circ  X)))
+dc\stackrel{\circ}{\wedge} d(a\circ(Div\;(b\circ  X))))$$
$$+6(da\stackrel{\circ}{\wedge}d(c\circ (Div\;(b\circ X)))
+db\stackrel{\circ}{\wedge}d(a\circ (Div\;(c\circ X)))
+dc\stackrel{\circ}{\wedge}d(b\circ (Div\;(a\circ X))))
$$
$$- 5 (da\stackrel{\circ}{\wedge} d(Div\;([b,c]\circ X))
+db\stackrel{\circ}{\wedge} d(Div\;([c,a]\circ X))
+dc\stackrel{\circ}{\wedge} d(Div\;([a,b]\circ X)))
$$
$$-d[a,b]\stackrel{\circ}{\wedge} d(Div\;(c\circ X))
-d[b,c]\stackrel{\circ}{\wedge} d(Div\;(a\circ X))
-d[c,a]\stackrel{\circ}{\wedge} d(Div\;(b\circ X))
$$
$$- 12 (d[a,b]\stackrel{\circ}{\wedge} d(c\circ Div\,X)
+d[b,c]\stackrel{\circ}{\wedge} d(a\circ Div\,X)
d[c,a]\stackrel{\circ}{\wedge} d(b\circ Div\,X))
$$
$$+ 27 d(a\circ [b,c]+b\circ[c,a]+c\circ[a,b])\stackrel{\circ}{\wedge}
d(Div\,X)$$
$$+ 18 Div\,a\; (db\stackrel{\circ}{\wedge}d(c\circ Div\,X)-
dc\stackrel{\circ}{\wedge}d(b\circ Div\,X))$$
$$+18 Div\,b \;( dc\stackrel{\circ}{\wedge}d(a\circ Div\,X)-
db\stackrel{\circ}{\wedge}d(a\circ Div\,X))
$$
$$+
18 Div\,c \;(da\stackrel{\circ}{\wedge}d(b\circ Div\,X)-
db\stackrel{\circ}{\wedge}d(a\circ Div\,X))
$$
$$- 14  Div\,a \;(db\stackrel{\circ}{\wedge} d(Div\;(c\circ X))-
dc\stackrel{\circ}{\wedge} d(Div\;(b\circ X)))$$
 $$-14 Div\,b\;(dc\stackrel{\circ}{\wedge} d(Div\;(a\circ X))-
da\stackrel{\circ}{\wedge} d(Div\;(c\circ X)))$$
$$-14  Div\,c\; (da\stackrel{\circ}{\wedge} d(Div\;(b\circ X))-
db\stackrel{\circ}{\wedge} d(Div\;(a\circ X)))
$$
\end{th}

We are not able to write escort invariant of type $(3,5,2)$ in a
compact form.

\begin{crl} $10$-commutator on $L=Vect_0(3)$
is well defined and escort invariants are given by
$$escort_{271}: R(\pi_1)\otimes \wedge^2 R(\pi_1+\pi_2)\otimes R(2\pi_1+\pi_2)
\rightarrow \wedge^2 R(\pi_1),$$

$$escort_{271}(u,a,b,X)=$$

$$du\wedge (11 d(Div\;([a,b]\circ X))+21 d(a\circ Div(b\circ X)-
b\circ Div(a\circ X)))$$
$$-32 (d(a\circ u)\wedge d(Div(b\circ X))-d(b\circ u)\wedge d(Div(a\circ X)))$$
$$+12 (da\stackrel{\circ}{\wedge}d((b\circ X)\circ u)-
db\stackrel{\circ}{\wedge}d((a\circ X)\circ u))$$
$$-28 d[a,b]\stackrel{\circ}{\wedge} d(X\circ u)$$
$$+16 (da\stackrel{\circ}{\wedge}d (X\circ(b\circ u))-
db\stackrel{\circ}{\wedge} d(X\circ (a\circ u))),$$

\bigskip

$$escort_{3601}: \wedge^3 R(\pi_1+\pi_2)\otimes R(3\pi_1+\pi_{n-1})\rightarrow
\wedge^2R(\pi_{1})$$

$$escort_{3601}(a,b,c,X)=$$
$$-6(da\stackrel{\circ}{\wedge} d(b\circ(Div\;c\circ  X))
+db\stackrel{\circ}{\wedge} d(c\circ(Div\;(a\circ  X)))
+dc\stackrel{\circ}{\wedge} d(a\circ(Div\;(b\circ  X))))$$
$$+6(da\stackrel{\circ}{\wedge}d(c\circ (Div\;(b\circ X)))
+db\stackrel{\circ}{\wedge}d(a\circ (Div\;(c\circ X)))
+dc\stackrel{\circ}{\wedge}d(b\circ (Div\;(a\circ X))))
$$
$$- 5 (da\stackrel{\circ}{\wedge} d(Div\;([b,c]\circ X))
+db\stackrel{\circ}{\wedge} d(Div\;([c,a]\circ X))
+dc\stackrel{\circ}{\wedge} d(Div\;([a,b]\circ X)))
$$

\bigskip

$$escort_{352}:\wedge^4R(\pi_1+\pi_2)\otimes\wedge^2R(2\pi_1+\pi_2)
\rightarrow \wedge^2R(\pi_1)\cong L_{-1}$$
is too big to be presented here.
\end{crl}

All coefficients are calculated by Mathematica and Maple. I am gratefull
to Y.I. Manin and M. Jibladze for interest to 10-and 13-~commutators.

\end{document}